\documentclass[bibyear]{aa}
\usepackage{graphicx}
\usepackage{color}
\usepackage{txfonts}\usepackage{subcaption}
\usepackage{natbib}
\usepackage{graphicx}
\usepackage{txfonts}
\usepackage{newtxtext,newtxmath}
\usepackage{ae,aecompl}
\usepackage{amsmath}    
\usepackage{amssymb}    
\usepackage{float}
\usepackage{multirow}
\everymath{\displaystyle}
\begin{document}

   \title{Relations between phenomenological and physical parameters
in the hot coronae of AGNs computed with the \textit{MoCA} code}
\titlerunning{AGNs hot coronae parameters with the 
        \textit{MoCA} code}

   \subtitle{}

  \author{R. Middei
          \inst{1}\fnmsep\thanks{riccardo.middei@uniroma3.it}
          \and
          S. Bianchi\inst{1} \and A. Marinucci \inst{1,2} \and G. Matt\inst{1} \and P.-O. Petrucci \inst{3}  \and F. Tamborra \inst{4} \and A. Tortosa \inst{5}
              }
\institute{Dipartimento di Matematica e Fisica, Universit\`a degli Studi Roma Tre, via della Vasca Navale 84, I-00146 Roma, Italy.
\and
ASI - Unit\`a di Ricerca Scientifica , Via del Politecnico snc, I-00133, Roma, Italy.
\and
Univ. Grenoble Alpes, CNRS, IPAG, F-38000 Grenoble, France.
\and
Nicolaus Copernicus Astronomical Center of the Polish Academy of Sciences, ul. Bartycka 18, 00716 Warsaw, Poland.
\and
INAF/Istituto di Astrofisica e Planetologia Spaziali, via Fosso del Cavaliere, 00133 Roma, Italy.
}



\abstract
{The primary X-ray emission in active galactic nuclei (AGNs) is widely believed to be due to Comptonisation of the thermal radiation from the accretion disc in a corona of hot electrons. The resulting spectra can, in first approximation, be modelled with a cut-off power law, the photon index and the high-energy roll-over encoding information on the physical properties of the X-ray-emitting region. The photon index and the high-energy curvature of AGNs ($\Gamma$, E$_c$) have been largely studied since the launch of X-ray satellites operating above 10 keV. However, high-precision measurements of these two observables have only been obtained in recent years thanks to the unprecedented sensitivity of \textit{NuSTAR} up to 79 keV.}
{We aim at deriving relations between $\Gamma$, E$_c$ phenomenological parameters and the intrinsic properties of the X-ray-emitting region (the hot corona), namely the optical depth and temperature.}
{We use \textit{MoCA} (Monte Carlo code for Comptonisation in Astrophysics) to produce synthetic spectra for the case of an AGN with M$_{\rm{BH}}$=1.5$\times$10$^8$ M$_\odot$ and $\dot m$=0.1 and then compared them with the widely used power-law model with an exponential high-energy cutoff.}
{We provide phenomenological relations relating $\Gamma$ and E$_{\rm{c}}$ with the opacity and temperature of the coronal electrons  for the case of spherical and slab-like coronae. These relations give origin to a well defined parameter space which fully contains the observed values. Exploiting the increasing number of high-energy cut-offs quoted in the literature, we report on the comparison of physical quantities obtained using \textit{MoCA} with those estimated using commonly adopted spectral Comptonisation models. Finally, we discuss the negligible impact of different black hole masses and accretion rates on the inferred relations.}
{}


\maketitle
%

\section{Introduction}
X-rays emerging from active galactic nuclei (AGNs) are the result of an inverse-Compton process occurring in the proximity of the central black hole (BH), where optical-UV photons arising from the accretion disk are inverse-Compton scattered by hot electrons in an optically thin, compact corona \cite[e.g.][for details on the two-phase model]{haar91,Haar93,haar94}. Such a Comptonisation mechanism accounts for the power-law-like shape of the X-ray primary continuum emission and the high-energy roll-over observed in various nearby AGNs \citep[e.g.][]{Nica00,Pero02,DeRo02,Molina09,Molina13,Mali14,Ricci18}. Broadband X-ray spectral investigations are of primary importance in studying AGN Comptonisation properties. Indeed, as extensively discussed in the literature \citep[e.g.][]{Ghisellini13} both the photon index ($\Gamma$) and high-energy cut-off ($E_{\rm{c}}$) of the X-ray primary emission depend on the intrinsic properties of the Comptonising medium, namely its temperature, optical depth, and geometry. Therefore, the interplay between coronal parameters and the AGN X-ray spectral shape has been the object of several investigations, especially with observatories capable of detecting hard X-rays. \citet{Dadina07}, using \textit{BeppoSAX} data, collected and studied the photon index and E$_{\rm{c}}$ of a sample of AGNs \citep[see also][for previous results]{Pero02}, while similar works were performed on \textit{INTEGRAL} data \citep[e.g.][]{Bassani06,Molina09,Mali14}, and, in the context of the BAT AGN Spectroscopic Survey, by \citet{Ricci18}.
Subsequently, \textit{NuSTAR} \citep{Harr13}, thanks to its unprecedented effective area above 10 keV, greatly helped in studying the exponential cut-offs of the nuclear continuum in several AGNs \citep[see e.g.][]{Fabi15,Fabi17,Tort18}. These space missions gave rise to a substantial corpus of high-energy cut-off and photon index measurements.\\
\indent While several Comptonisation models exist in the literature, many of them also available for direct fitting to the data, they are usually valid in a limited range of parameters. Moreover, in some papers, especially in the past and/or when a large sample of sources were analysed, only the phenomenological parameters are provided. It could therefore be of some help to have relations 
converting phenomenological parameters to physical ones. \citet{Petr00} provided an approximate relation between the observed high-energy cut-off of NGC 5548 and the thermal energy of the corona (E$_{c}\sim$2-3kT), while \cite{Petr01} discussed this for a larger sample of Seyfert 1 galaxies observed with \textit{BeppoSAX}. However, this relation has proven to be accurate 
only for an extended slab geometry. On the other hand, \cite{Belo99} reports a relation between the spectral photon index and the Compton parameter $y$. This parameter is the product of the average fractional energy change per scattering and the mean number of scatters that the photon undergoes. Subsequently, $y$ is used to characterise the energy gain of a photon scattering within a finite medium. The Compton parameter encodes the physical conditions of the coronal plasma, and it can be defined as follows:
\begin{equation}
y=4(\theta+4~\theta^2)\tau(\tau+1)~,
\end{equation}
where $\theta=kT/m_{e}c^2$, and $\tau$ is the Comptonising medium opacity. Furthermore, \citet{Belo99} found that the photon index is related to $y$ according to $\Gamma\approx\frac{4}{9}~y^{-2/9}$.\\
\indent In this paper, we derive and discuss formulae which directly relate the phenomenological quantities $\Gamma$-E$_c$ with the corresponding coronal properties $kT$-$\tau$ for the cases of a slab-like and spherical corona. These relations were obtained using \textit{MoCA} \citep[Monte Carlo code for Comptonisation in Astrophysics,][]{Tamb18}.
\section{\textit{MoCA} simulations and setup}
\subsection{Hot corona and accretion disc}
\indent  Comptonisation Monte Carlo code (\textit{MoCA}; see \citet{Tamb18} for a detailed description of the code) is based on a single photon approach,  working in a fully special relativistic scenario. \textit{MoCA} allows for various and different physical and geometrical conditions of the accretion disc and of the Comptonising corona. In this paper, the corona is assumed to have either a spherical or a slab-like geometry, and to be as extended as the disc, whose radii have been set to be R$_{out}=500~ $ r$_{g}$ and  R$_{in}=6$ r$_{g}$, respectively.
\begin{figure}[h!]
        \includegraphics[width=0.48\textwidth]{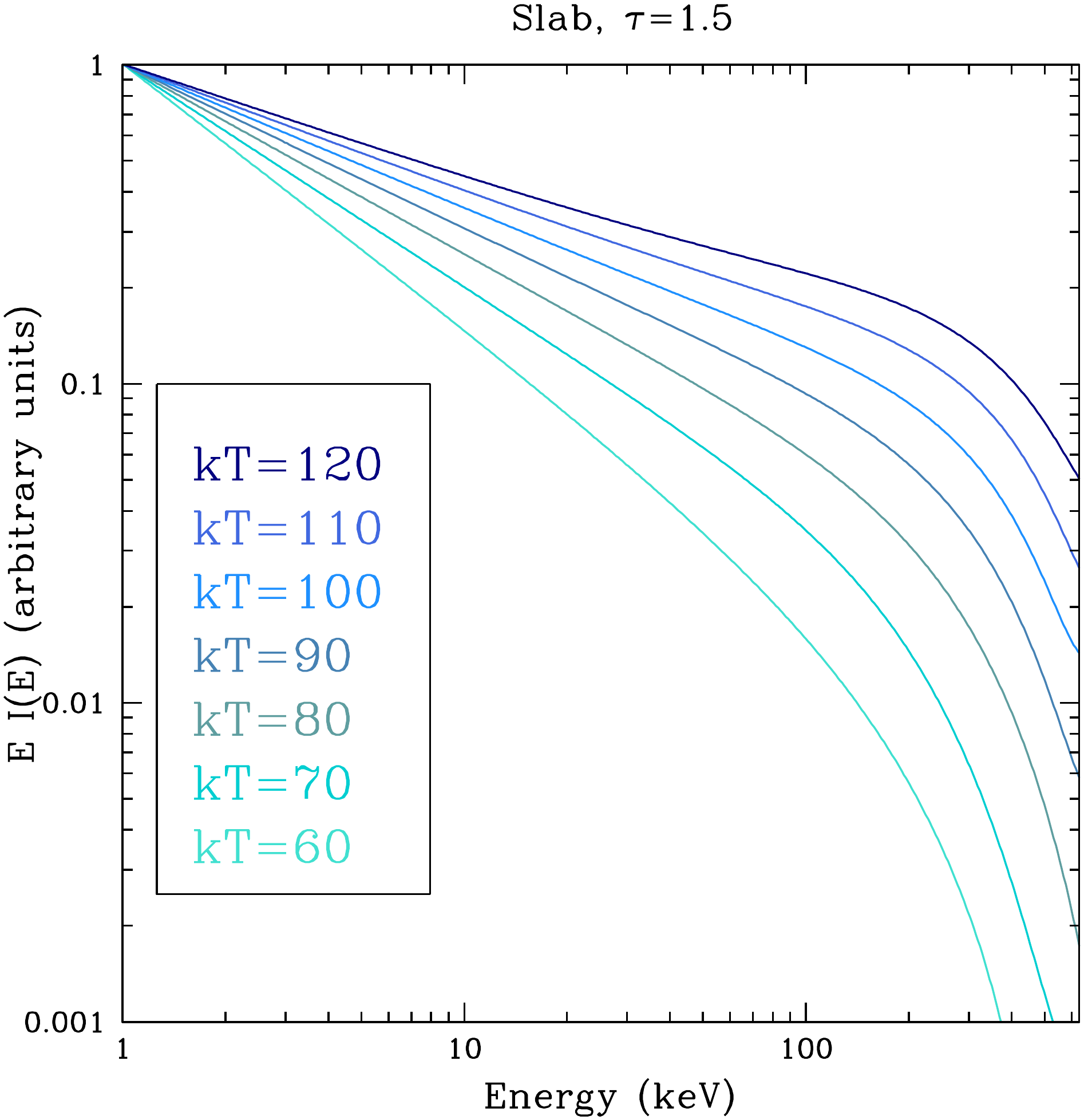}  
        \includegraphics[width=0.48\textwidth]{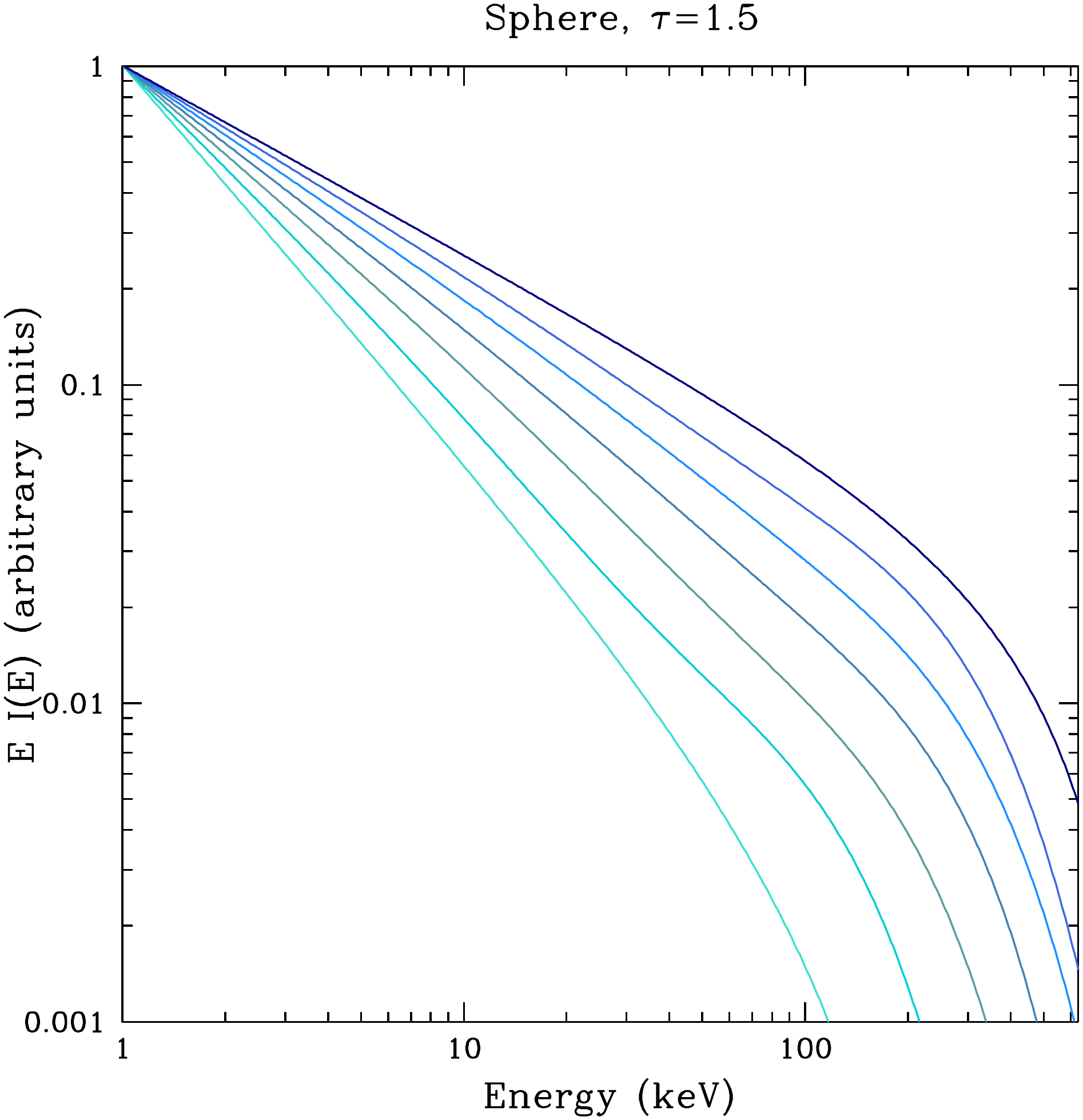}
        \caption{\small{\label{smspec} The $E~ I(E)$ graphs of the \textit{MoCA} simulation normalised at 1 keV. Curves refer to a fixed $\tau$=1.5 and different electron temperatures. The colour code is the same for the two panels. It is worth noting that the same $\tau$-$kT$ couple gives rise to different spectra when the slab-like or the spherical geometries are considered.}}
\end{figure}
Even though arguments (e.g. variability, \citep[][and references therein]{Uttley14}, microlensing \citep[][]{Chartas09,Morgan12} and timing \citep{Kara16,DeMa13}) exist that favour a compact corona, we used extended coronae. In fact, as discussed by \citet{Mari19}, Comptonised spectra emerging from compact corona (R$_{out}=100$ r$_{g}$-R$_{in}=6$) do not deviate significantly from those produced in more extended corona; see their Figure 3. The adoption of even more compact coronae (R$_{out}=20$ r$_{g}$, R$_{in}=6$) results only in the need for higher optical depths to recover the same spectral shape for a given temperature. However, in such cases, general relativity (GR) effects are not negligible \citep[see][for a detailed discussion on this topic]{Tamb18}, and the present version of \textit{MoCA} does not include GR.
For the slab-like geometry case, \textit{MoCA} allows the user to set up the corona height above the accretion disc (set to $10~r_{g}$ in our simulations). We use synthetic spectra computed assuming the source BH mass and accretion rate to be the same as those of \textit{Ark 120} \citep[e.g.][ and references therein]{Mari19}, namely $M_{BH}=1.5\times10^8M_\odot$ and $\dot m=L_{bol}/L_{Edd}=0.1$. For both the slab and spherical hot electron configurations, we simulated the Comptonised spectra using a wide range of values for electron temperature and  optical depth: $0.1\textless\tau\textless7$ and $20\textless$kT$\textless200$ keV, and in Fig.~\ref{smspec} we show a sample of spectra obtained by \textit{MoCA}. Moreover, spectra are computed from 0.01 keV up to 700 keV using 1,000 logarithmic energy bins, and a Poissonian error accompanies each spectral point.
The obtained spectra are averaged over the inclination angle and in Fig.~\ref{smspec} we show some exemplificative spectra normalised at 1 keV accounting for the two geometries considered in this work.\\

\subsection{Analysis}
\begin{figure*}
        \includegraphics[width=0.48\textwidth]{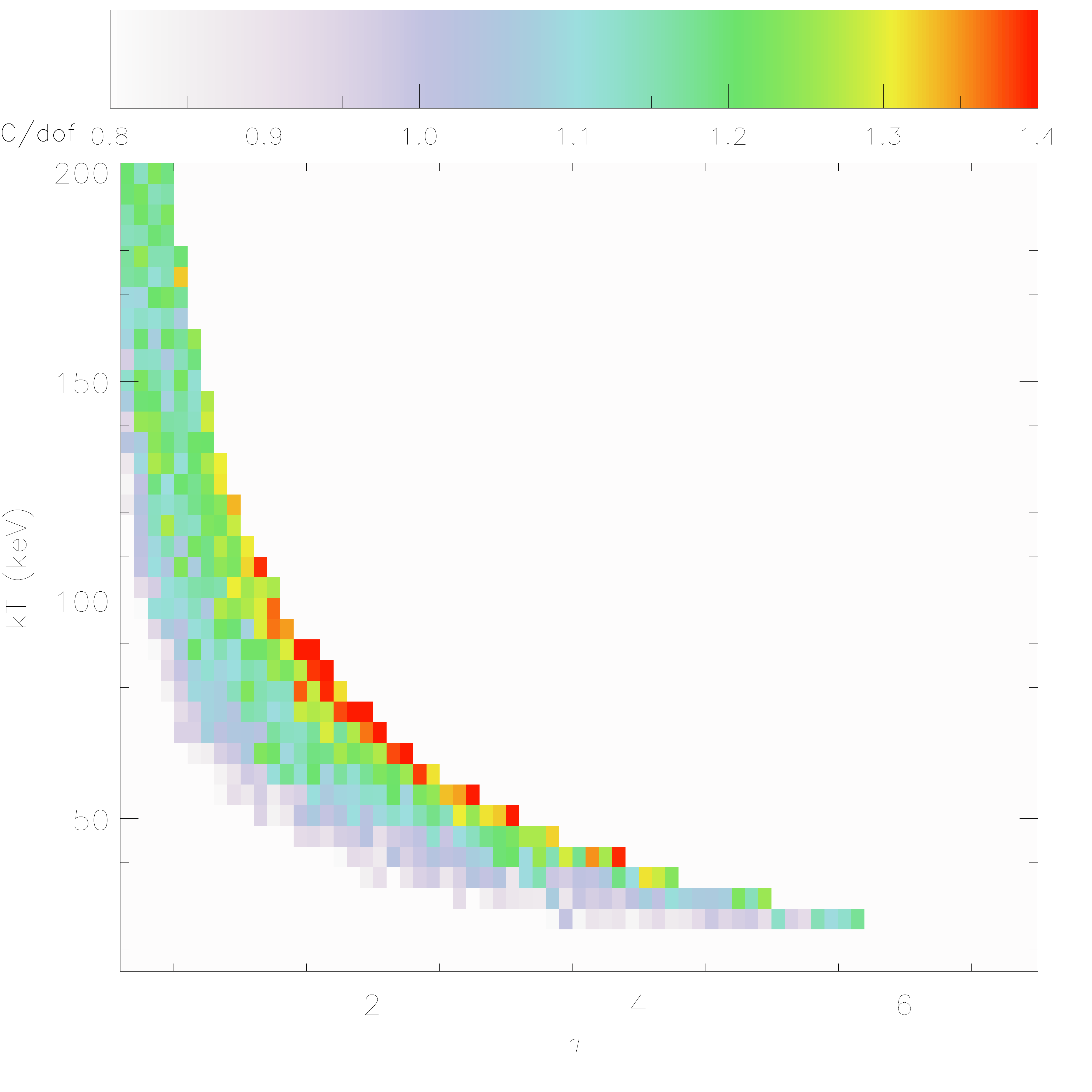}
        \includegraphics[width=0.48\textwidth]{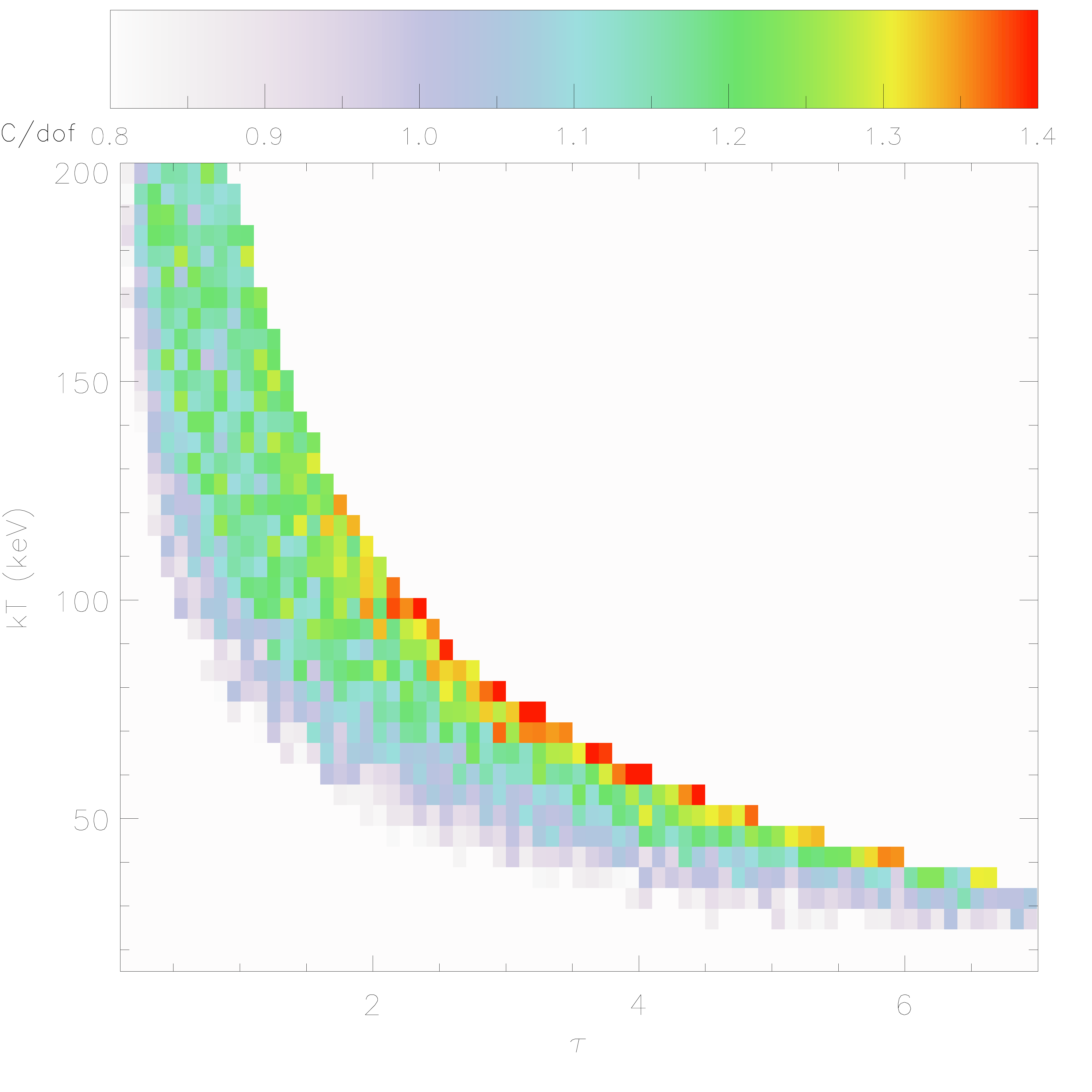}   
        \caption{\small{\label{maps} Map of the ratios between the $C$ statistics and the degrees of freedom for the slab-like (left) and spherical (right) cases.} The cut-off power law is a good approximation of the simulated spectra.}
\end{figure*}

\begin{figure}
        \includegraphics[width=0.48\textwidth]{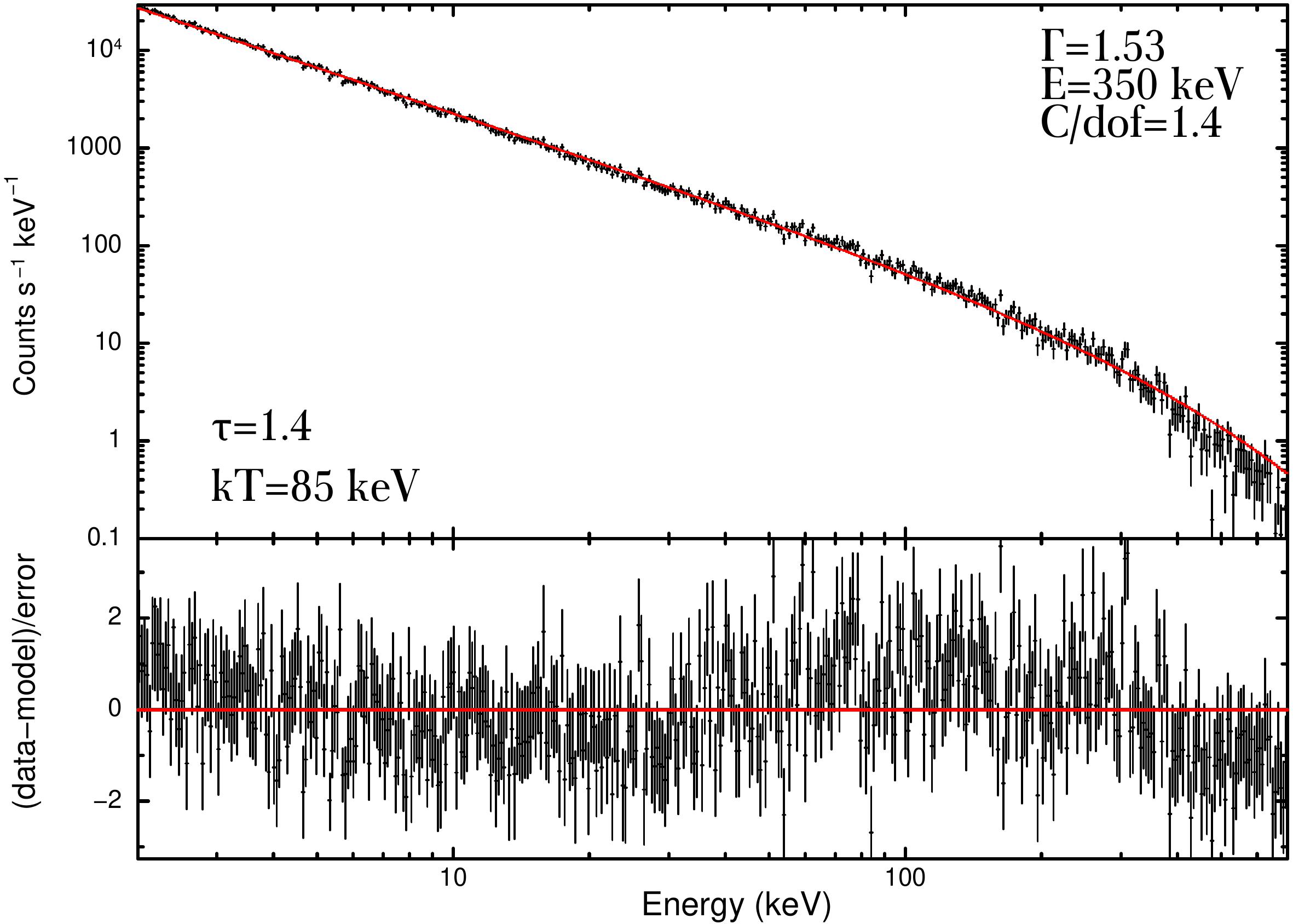}
        \caption{\small{Case of a spectrum with a relatively high $C$/d.o.f. emerging from a slab corona with $\tau=1.4$ and kT=85 keV. The cut-off power law leaves residuals mainly above 100 keV\label{bad}.}}
\end{figure}

\indent To find relations between the phenomenological and physical coronal parameters, we fitted the simulated spectra with a power law with an exponential cut-off \citep[model {\it cutoffpl} in \text{Xspec},][]{Arna96}. During the fitting procedure, the high-energy cut-off, the normalisation, and the photon index of the primary continuum emission are free to vary. Fits are performed in the 2-700 keV energy interval, or up to the last populated energy bin, if lower than 700 keV. In the fits, the \textit{C} statistic has been used \citep{Cash79}. These steps yield a corresponding fitted $\Gamma$-E$_c$ pair for each $kT$-$\tau$ couple. At this stage of the analysis, we excluded all the spectra for which we obtained a photon index outside the interval 1.5-2.5. In fact, we have no evidence of AGNs with such a flat or steep photon index \citep[e.g.][]{Bian09,Sobolewska09,Serafinelli17}.
In general, the cut-off power law is a very good approximation of the simulated spectra and in Fig.~\ref{maps} we show the goodness of the fits (quantified in terms of the $C$/d.o.f. ratio) as a function of the coronal properties. We notice that this ratio increases towards those regions corresponding to photon indices as flat as $\sim$1.5 and residuals with respect the cut-off power law mainly occur at high energy; see Fig.~\ref{bad}. These high-energy residuals cannot be observed with current facilities; indeed, they are outside the bandpass of the major operating observatories, and when instruments are sensible to such energetic photons these deviations are not detectable with the available signal-to-noise ratio.  


\begin{figure*}[h]
        \centering
        \includegraphics[width=0.48\textwidth]{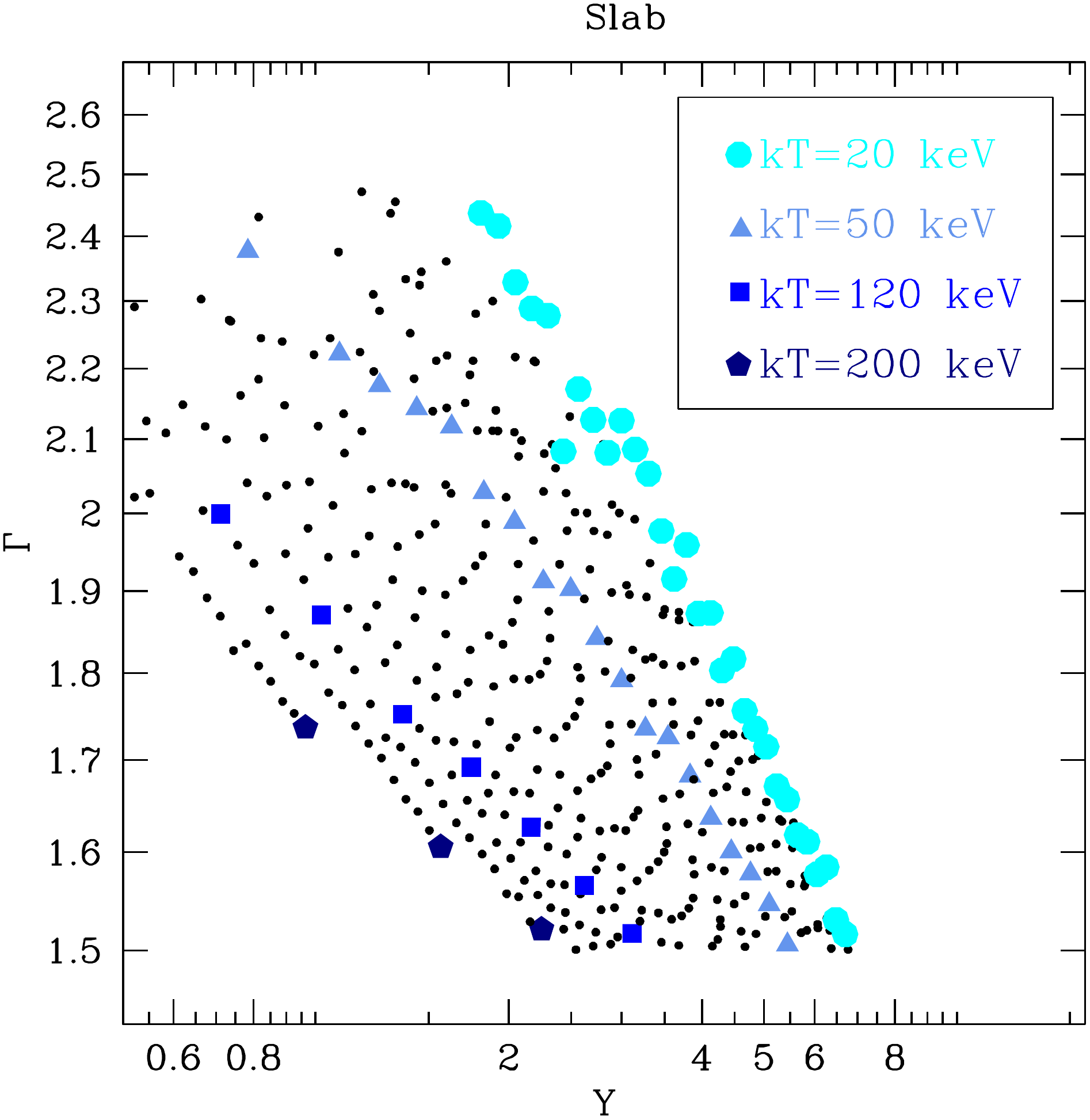}
        \includegraphics[width=0.48\textwidth]{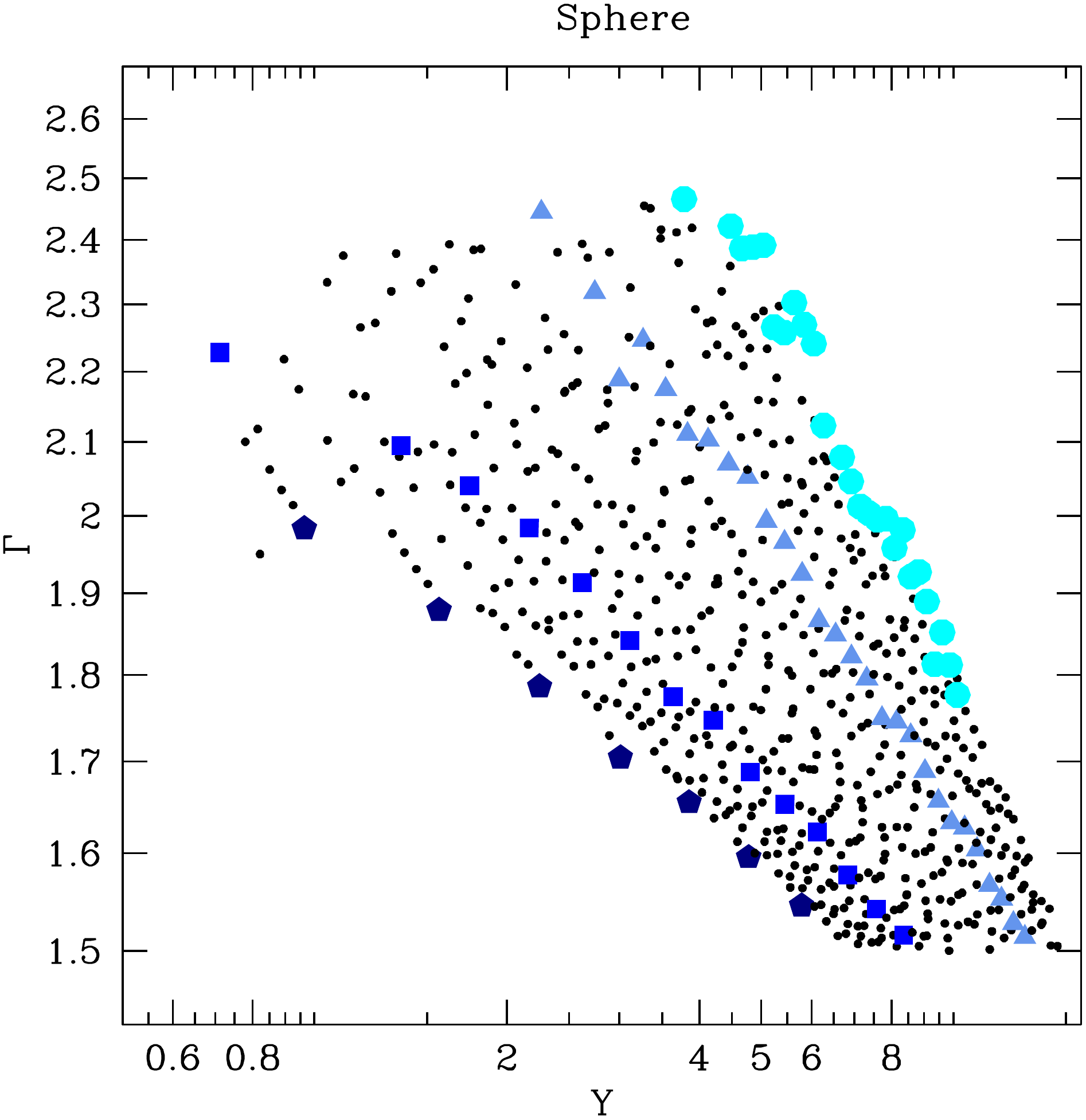}        
        \caption{\small{\label{fbag} Photon index as a function of the Compton parameter. The $y$ parameter is computed for the various optical depths and for all the temperatures (kept fixed). The colour code is the same for both panels.}}
\end{figure*}
\begin{figure*}[h]
        \centering
        \includegraphics[width=0.48\textwidth]{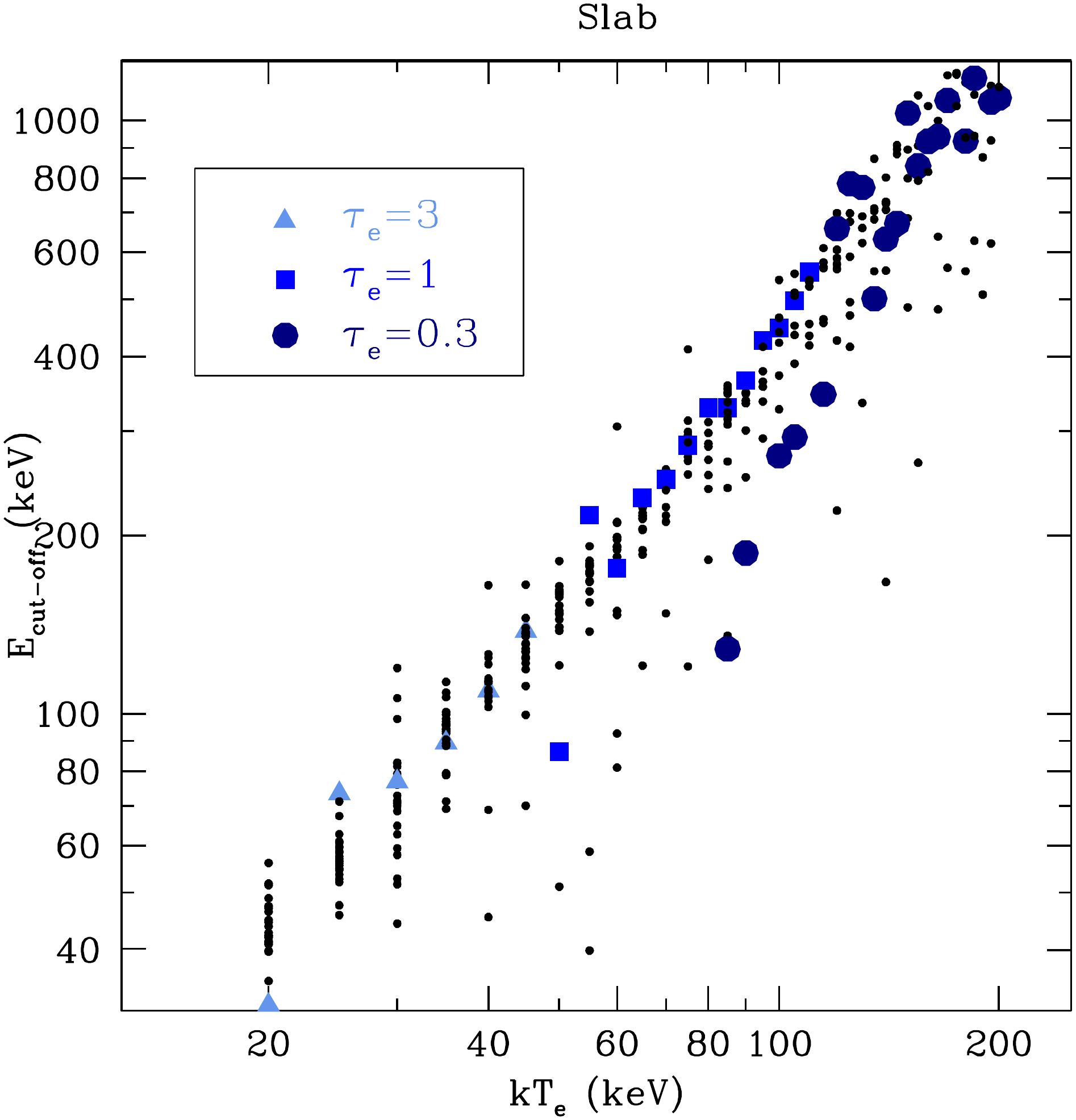}
        \includegraphics[width=0.48\textwidth]{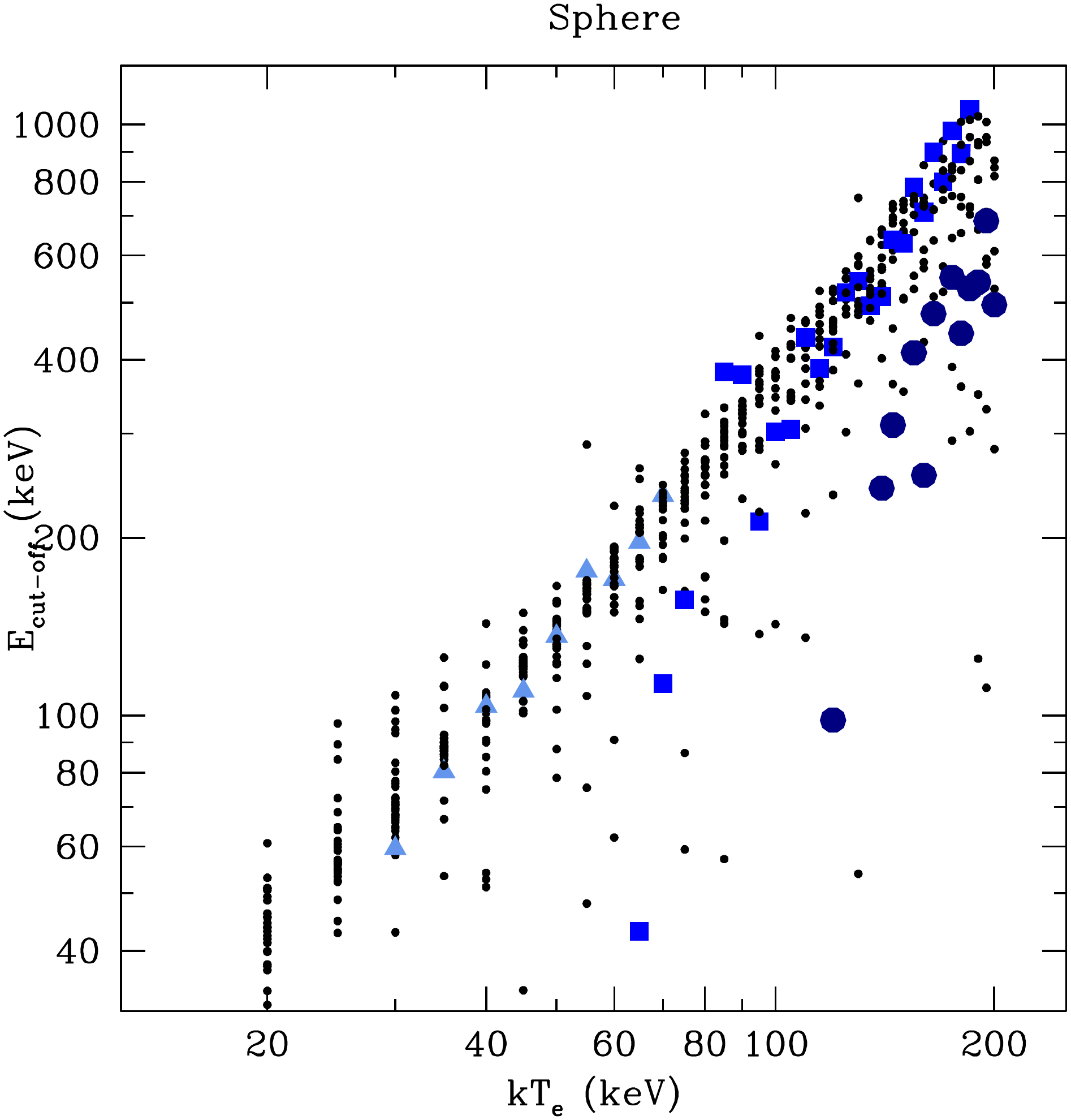}
        \caption{\small{\label{fbae} High-energy cut-off as a function of the electron temperature, for a fixed optical depth. Colours are the same in both graphs.}}
\end{figure*}

To derive relations connecting $\Gamma$-E$_c$ with $kT$-$\tau$, we started analysing the dependence of the photon index on the Compton parameter $y$. For each $\Gamma$-E$_{\rm{c}}$ couple, we then computed the corresponding $y$ using the proper $\tau-kT$. In Fig.~\ref{fbag}, we plot the photon index as a function of $y$ for different values of $\tau$ and for a fixed $kT$. We then performed log-log fits between $\Gamma$ and $y$. In the fits, both the normalisation and the slope are free to vary. These $\Gamma$ and the Compton parameter $y$ are clearly correlated, although points are widely dispersed; see  Fig~\ref{fbag}. On the other hand, we found that such a widespread dispersion is due to a further dependence of $\Gamma$ on the coronal temperature, which we quantified studying how the slopes ($a$) and normalisations ($b$) of the fits behave as a function of $kT$. We then performed additional fits between the following parameters: $a-kT$ and $b-kT$.
This procedure leads to Eq.~\ref{eqn:fbag} for the case of a slab-like corona:

\begin{align}
\label{eqn:fbag}
\begin{split}
\\
log \Gamma (\theta, \tau)=a(\theta) log~y (\theta, \tau)+b(\theta)
\\
a(\theta)=-0.34+1.02\times\theta-1.46\times \theta^2
\\
b(\theta)=0.47-1.02\times \theta+1.64\times \theta^2,
\end{split}
\end{align}
where $\theta=\frac{kT}{m_ec^2}$. For the spherical case we obtain
\begin{align}
\label{eqn:fsfg}
\begin{split}
\\
log \Gamma (\theta, \tau)=a(\theta) log~y (\theta, \tau)+b(\theta)
\\
a(\theta)=-0.35+0.87\times \theta-1.04\times \theta^2
\\
b(\theta)=0.60-1.53\times \theta+1.83\times \theta^2.
\end{split}
\end{align}

In accordance with these formulae, the photon index does not depend only on $y$, but is rather a function of $kT$ and $y$. To estimate the associated mean uncertainty $\Delta\Gamma/\Gamma$ for Eqs.~\ref{eqn:fbag} and~\ref{eqn:fsfg} we compared the inferred photon indices with values obtained fitting the simulations. The average of all the percentage errors yields a $\Delta\Gamma/\Gamma$ of $\sim$1.5\%(1\%) for the case of a slab-like (spherical) hot-corona. For the case of spherical geometry, we compared our Eq.\eqref{eqn:fsfg} with the relation by \cite{Belo99}: $\Gamma\approx\displaystyle\frac{4}{9}y^{-2/9}$. The author computed his equation by studying spectra simulated with the code by \cite{Coppi92} for only two values of $kT$ (50 and 100 keV) and leaving  the electron opacity free to vary. To estimate its uncertainty, we calculated the expected photon index using the relation by \cite{Belo99} for all the $kT$-$\tau$ couples. The comparison of these estimates with the fitted $\Gamma$ leads to an average percentage uncertainty of $\Delta\Gamma$/$\Gamma$=14\%, showing that this simplified formula significantly deviates from the \textit{MoCA} predictions.\\
\indent In a similar fashion, we derived relations connecting the high-energy cut-off with the physical properties of the Comptonising plasma.
The high-energy roll-over is mainly dependent on the coronal temperature. Thus, we started studying the $E_{\rm{c}}$ as a function of $kT$ keeping $\tau$ fixed; see Fig.~\ref{fbae}. The dependence of $E_{\rm{c}}$ on the plasma temperature is estimated by performing log-log fits between this parameter and $kT$ for different values of $\tau$. 
On the other hand, the obtained slopes ($\alpha$) and normalisations ($\beta$) are found to depend on the coronal opacity. Therefore, we quantified this dependence on the optical depth by performing linear fits for $\alpha$ and $\beta$ as a function of $\tau$. For the case of a slab-like corona these steps yield the following relations:
\begin{align}
\label{eqn:fbae}
\begin{split}
\\
log\frac{E_c}{keV}=\alpha(\tau)\times(log~\theta+2.71)+\beta(\tau)
\\
\alpha(\tau)=2.48-0.57\times\tau+0.065\times\tau^2
\\
\beta(\tau)=-2.33+1.16\times\tau-0.14\times\tau^2.
\end{split}
\end{align}
 For a spherical geometry we obtain
\begin{align}
\label{eqn:fsfe}
\begin{split}
\\
log\frac{E_c}{keV}=\alpha(\tau)\times(log~\theta+2.71)+\beta(\tau)
\\
\alpha(\tau)=2.98-0.66\times\tau+0.054\times\tau^2
\\
\beta(\tau)=3.35+1.3\times\tau-0.11\times\tau^2.
\end{split}
\end{align}

\begin{figure*}
        \centering
        \includegraphics[width=0.48\textwidth]{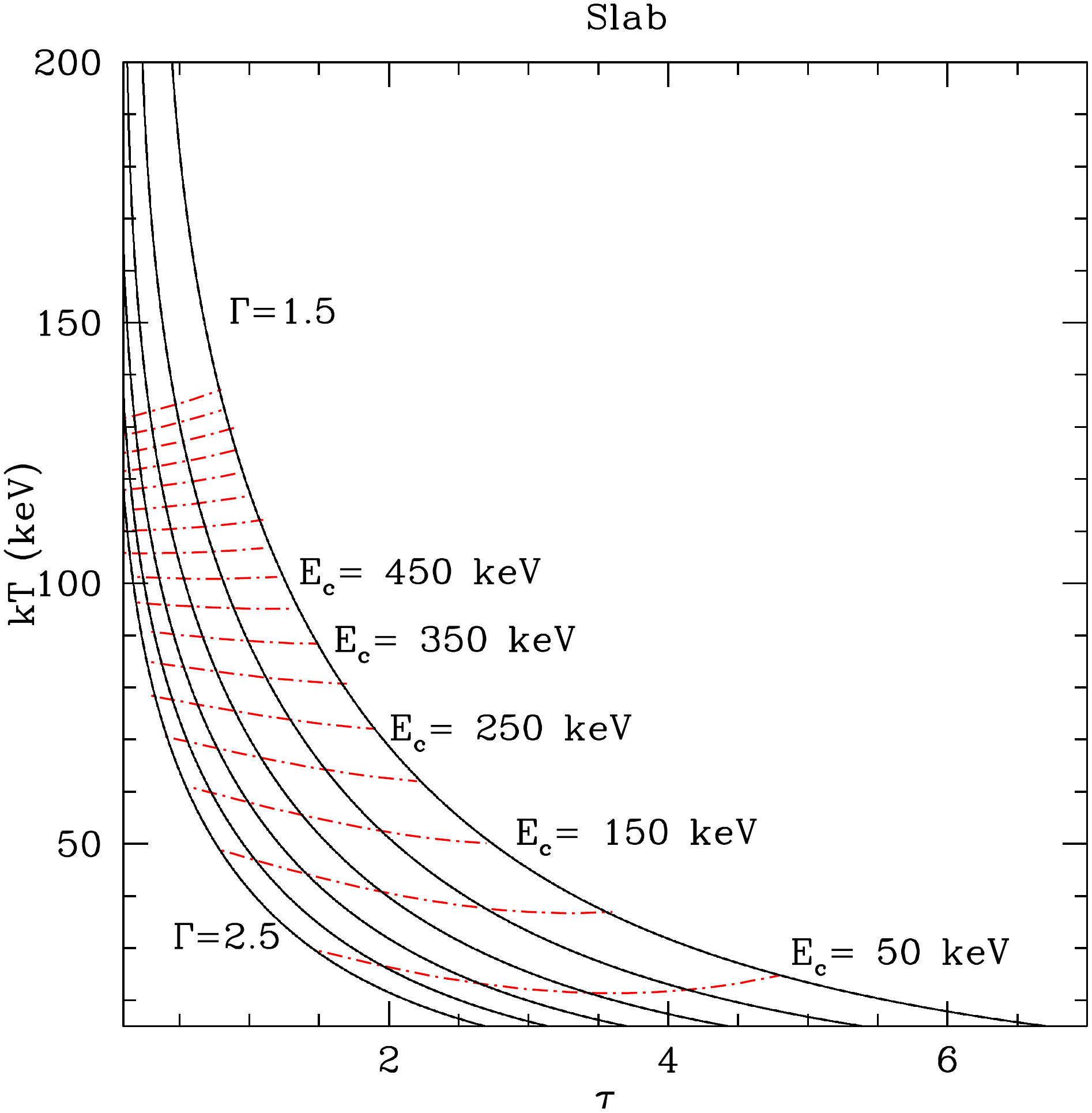}       
        \includegraphics[width=0.48\textwidth]{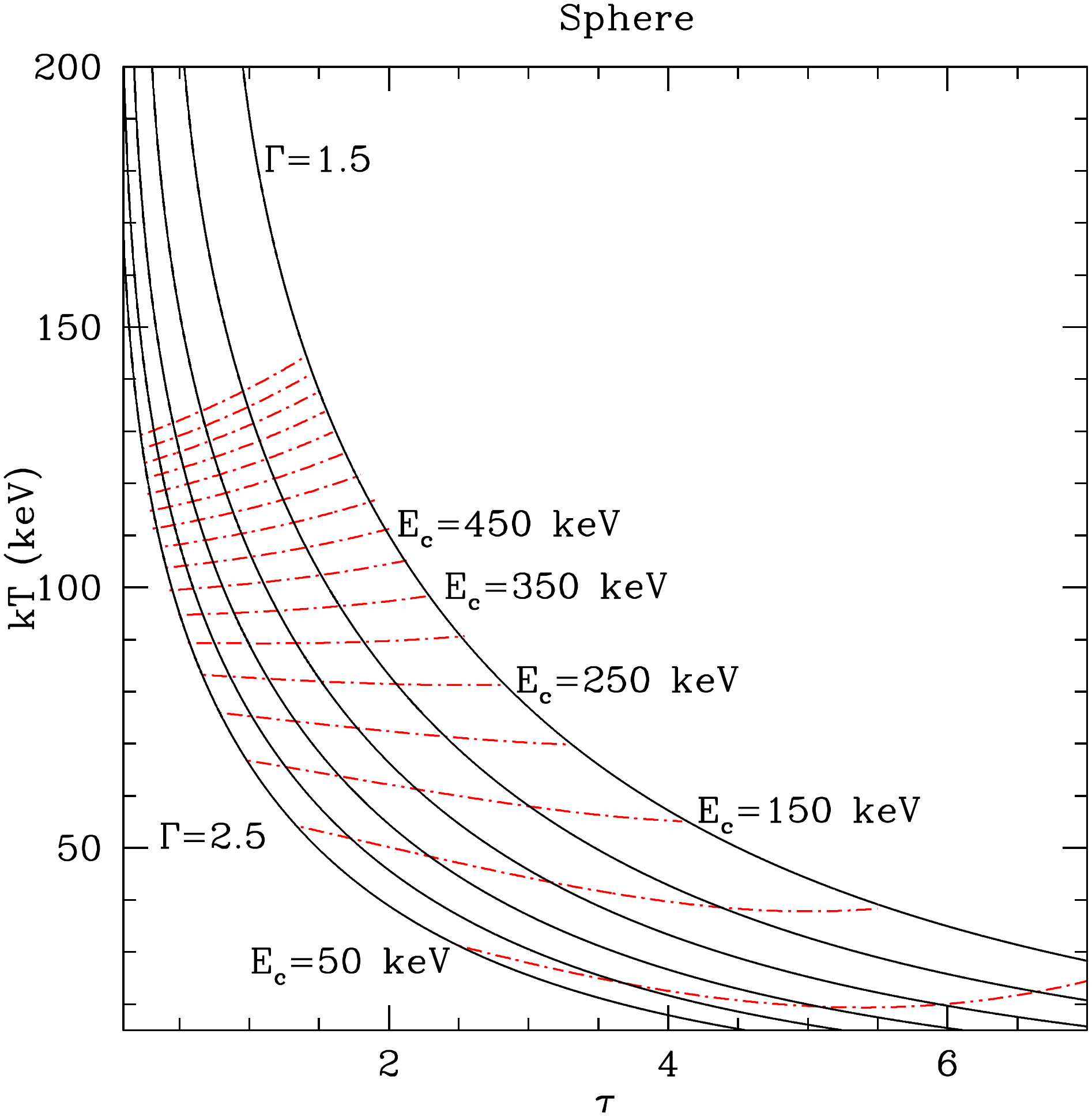}       
        \caption{\label{bagrid} Grids in the physical parameters space computed using Eqs.~\ref{eqn:fbag},~\ref{eqn:fsfg},~\ref{eqn:fbae}, and ~\ref{eqn:fsfe}. Black solid lines represent the iso-$\Gamma$ curves, while iso-$E_{\rm{c}}$  are shown via the red solid-dashed lines. Iso-$\Gamma$ curves have a 0.2 step and $\Gamma$ steepens from right to left. A 50 keV step is used to represent the various high-energy cut-off iso-curves.}
\end{figure*}
\begin{figure}
        \centering
        \includegraphics[width=0.48\textwidth]{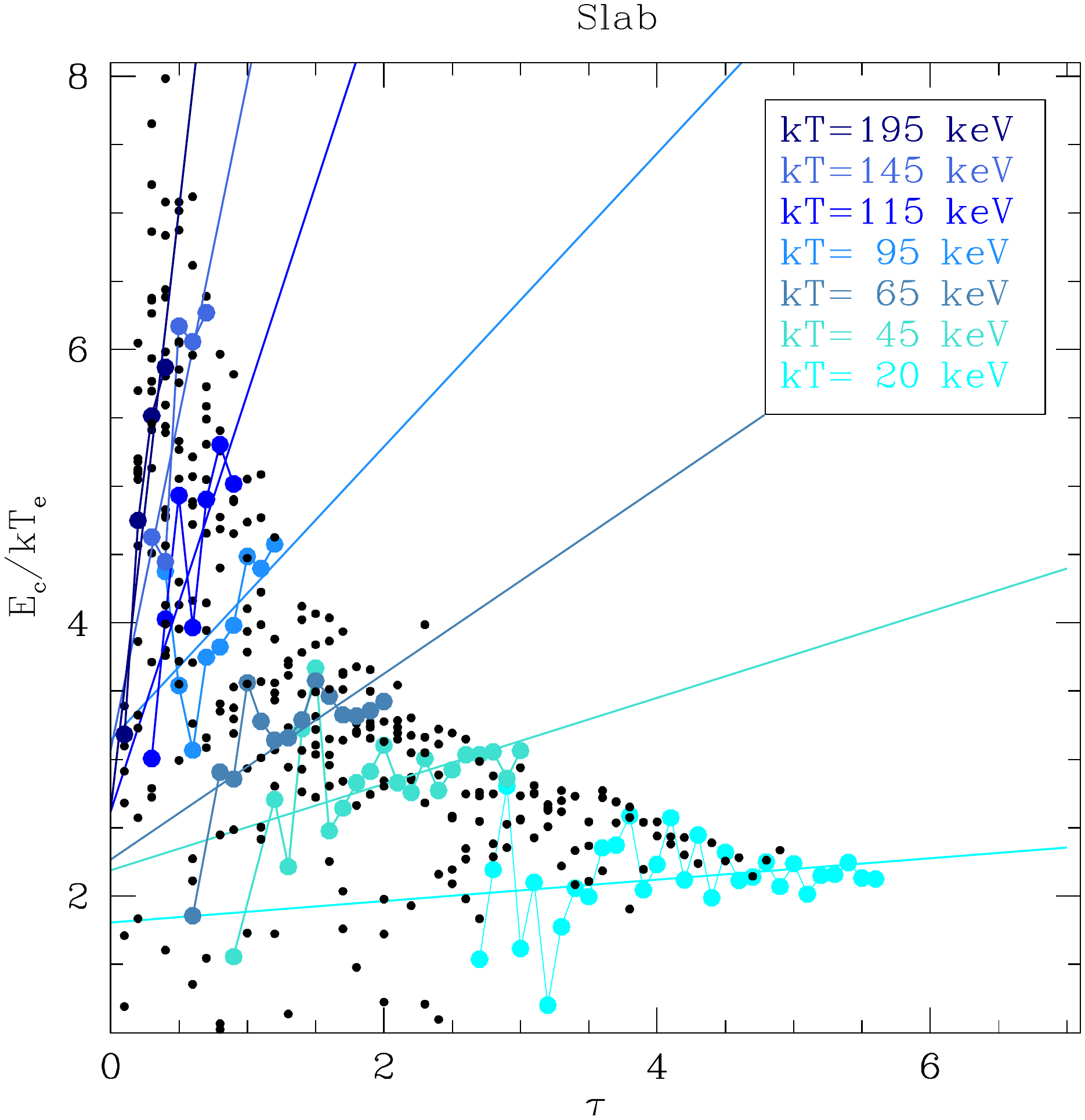}    \includegraphics[width=0.48\textwidth]{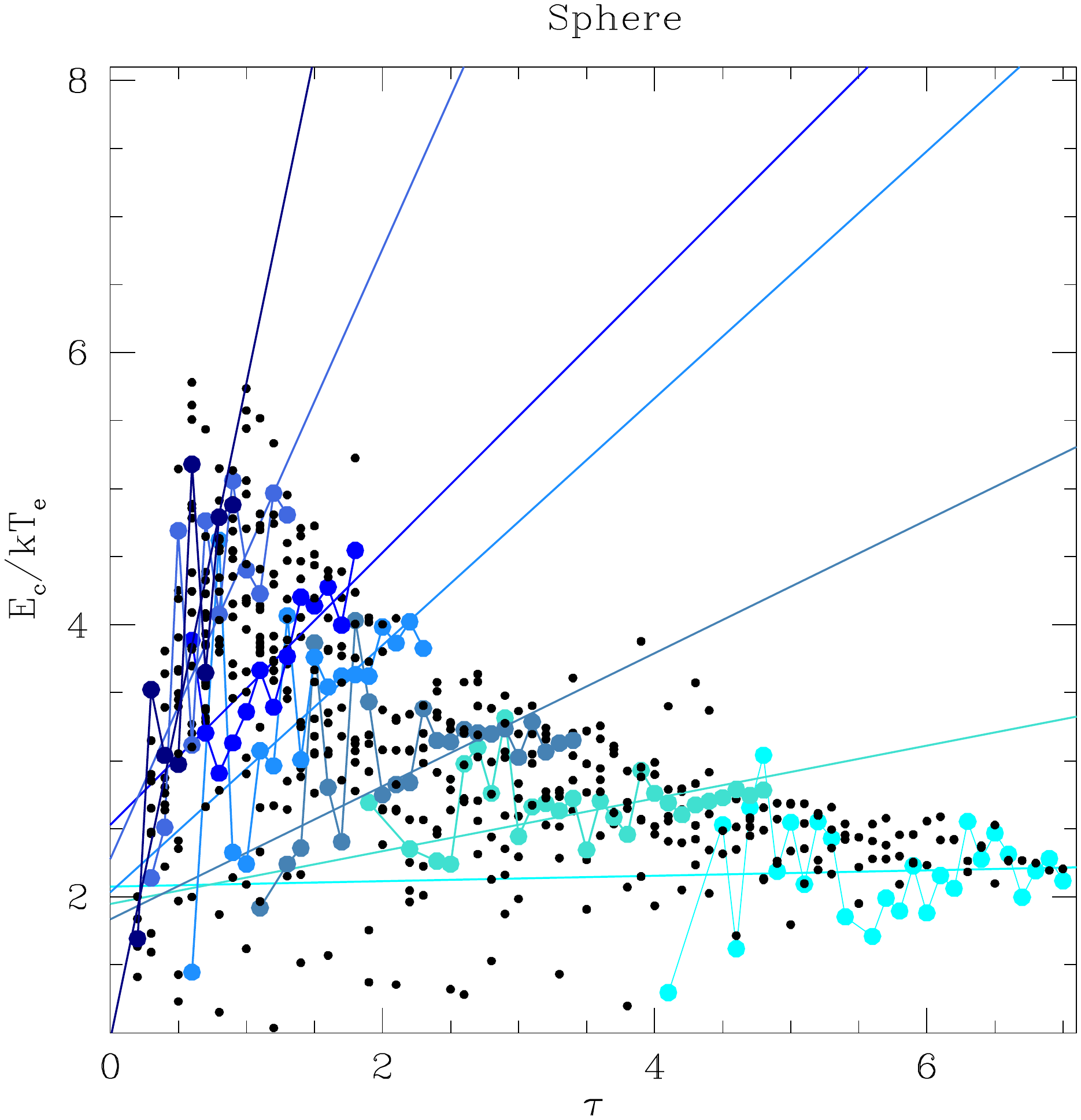}  
        \caption{\label{nopop} Ratio between the high-energy cut-off E$_{\rm{c}}$ and the coronal temperature $kT$  displayed as a function of the optical depth $\tau$ for both geometries. In both panels, straight solid lines represent the best-fit between E$_{\rm{c}}/kT$ and $\tau$ at a fixed temperature kT$_{\rm{e}}$, while lines connecting the points are reported for plot purposes only.}
\end{figure}
The uncertainty associated with Eq.~\ref{eqn:fbae} is $\Delta E_c/E_c$$\simeq$13\%, while for the case of a spherical corona we obtain $\Delta E_c/E_c\simeq$17\%. For both geometries considered here, the high-energy cut-off not only depends on the electron temperature, but also on the coronal opacity. The above equations directly relate the high-energy cut-off with the coronal temperature and encode the dependence on opacity of the Comptonising electrons.\\
\indent  Once the dependence between the phenomenological parameters and the physical $kT$ and $\tau$ have been quantified, we use Eqs.~\ref{eqn:fbag} and \ref{eqn:fbae} for the slab-like coronae and Eqs.~\ref{eqn:fsfg} and \ref{eqn:fsfe} for the spherical coronae to determine the interesting regions in the $kT$-$\tau$ parameters space. In particular, by plotting iso-$\Gamma$ curves for $1.5\leq\Gamma\leq2.5$ and iso-E$_c$ lines, we identify regions in the physical parameters plane corresponding to typically measured AGN photon indices and high-energy cut-offs; see Fig.\ref{bagrid}. On average, for a certain $\Gamma$, a slab corona returns lower values of optical depth and temperature, while to reach the same spectral shape, higher $kT$ and $\tau$ are required when a sphere-like corona is considered. Moreover, iso-E$_c$ curves behave differently for the two geometries. Indeed for the sphere, high-energy cut-offs generally correspond to higher temperatures ($\gtrsim$10 keV more) if compared with those measured assuming a slab corona. Furthermore, we observe that the commonly adopted relation $E_{\rm{c}}\sim2-3~kT$ is not valid for all the $kT$ and $\tau$ regimes; see Fig.~\ref{nopop}.

In this figure, we show the high-energy cut-off and electron temperature ratio as a function of the coronal opacity. In particular, we point out that for low-temperature regimes the optical depth does not play any role in determining the cut-off energy of the spectrum, while at larger electron temperatures, the E$_{\rm{c}}$/$kT$ ratio is no more insensitive to the coronal opacity.
\begin{figure*}
        \centering
                \includegraphics[width=0.48\textwidth]{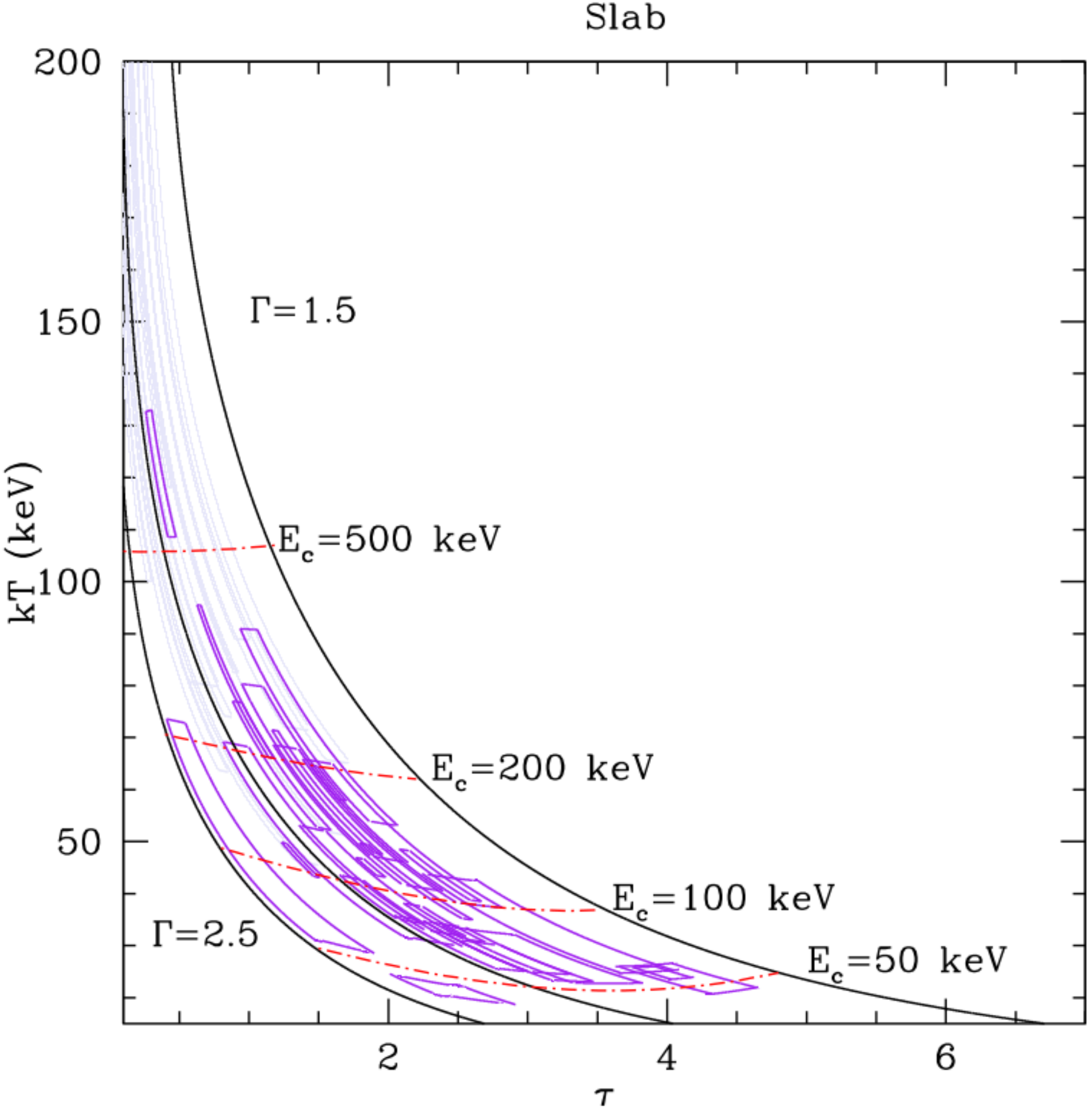}     
\includegraphics[width=0.48\textwidth]{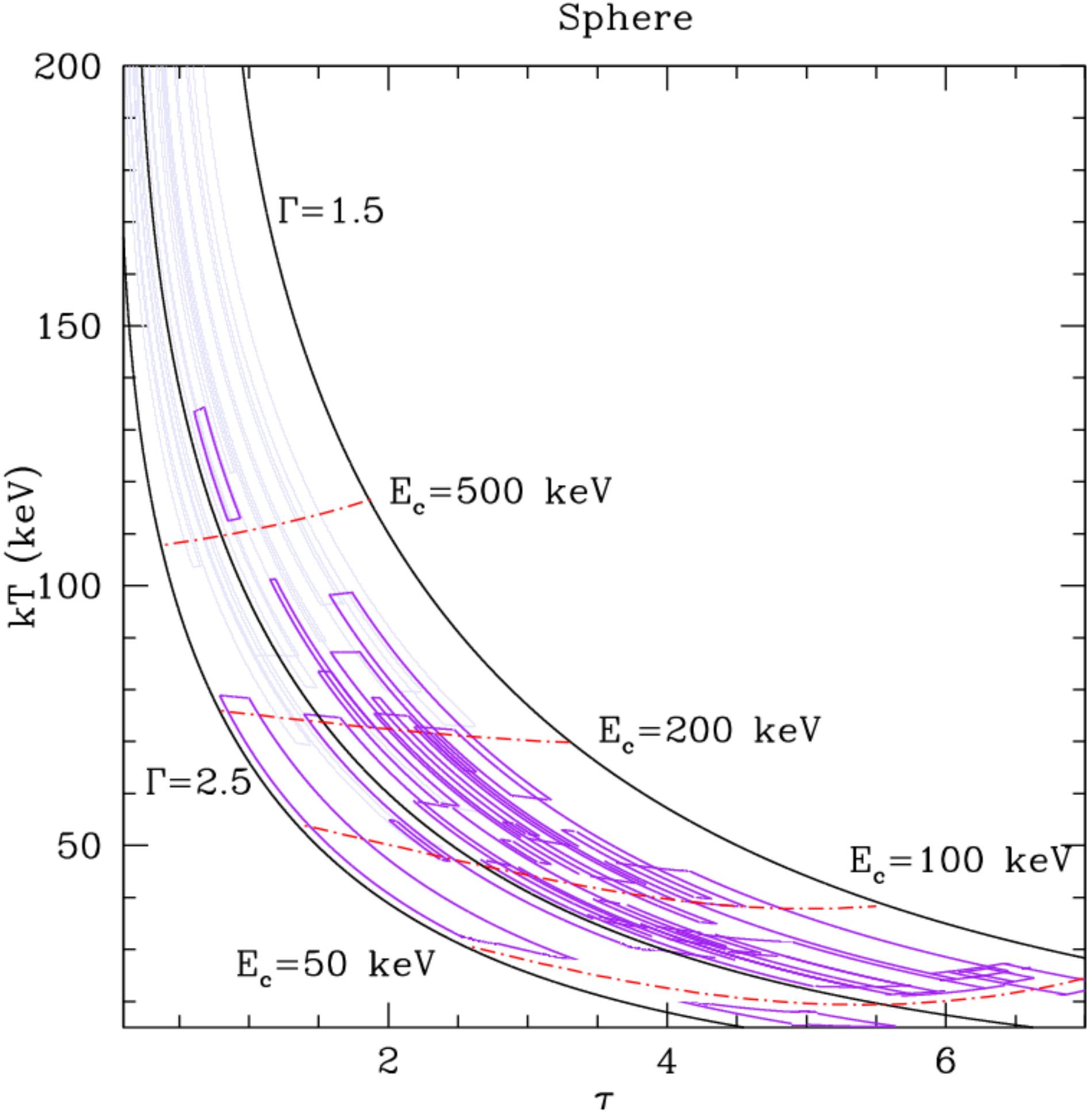}
        \caption{\label{bagrids}  Contours in the physical parameter space for all the AGNs reported in Table 1. Magenta contours correspond to measured high-energy cut-off while blue is used for AGNs characterised by an unconstrained high-energy cut-off. Moreover, black and red curves emphasise the iso-$\Gamma$ and iso-E$_{\rm{c}}$ labelled in the plot.}
\end{figure*}

Since a dependence of the Comptonised spectra on the spectral shape of the seed photons would affect our conversions and the derived coronal parameters, here we check if Eqs.~\ref{eqn:fbag},~\ref{eqn:fsfg},~\ref{eqn:fbae}, and ~\ref{eqn:fsfe} can be broadly used regardless of the AGN black hole mass and accretion rate. We computed simulations for M$_{\rm{BH}}$=10$^{7}$ M$_\odot$ and $\dot{m}$=0.1, M$_{\rm{BH}}$=10$^{8}$ M$_\odot$ with $\dot{m}$=0.05 and $\dot{m}$=0.2, and, finally, for M$_{\rm{BH}}$=10$^{9}$ M$_\odot$ and $\dot{m}$=0.1. Following the same procedure described in Sect. 3, we fitted these spectra using a cut-off power-law model within \textit{Xspec}. As in the previous case, this allowed us to build a new database of correspondences between the phenomenological and physical parameters. Neither the photon index nor the high-energy cut-off estimates are strongly dependent on the particular values of black hole masses and accretion rate. In fact, we notice that the fitted values for the phenomenological parameters are within the uncertainties associated to the equations reported in this section. 
\section{Estimates of coronal parameters in local AGNs}
\subsection{The \textit{NuSTAR} sample}
Our relations can be used for probing the coronal properties of AGNs. From  Eqs.~\ref{eqn:fbag}, \ref{eqn:fsfg}, \ref{eqn:fbae}, and \ref{eqn:fsfe}, we can directly obtain the coronal properties of those Seyfert galaxies for which $E_{\rm{c}}$ and $\Gamma$ are available. Here, we consider the AGN sample presented by \cite{Tort18}, which we updated to include recently observed AGNs. Moreover, when multiple values are found in the literature for the same source, $\Gamma$ and $E_c$ measurements based on broadband analyses have been preferred. The use of our formulae on the chosen phenomenological parameters $\Gamma$ and E$_{\rm{c}}$ yielded Table \ref{newtort}, in which we report the \textit{MoCA}-based estimates for the coronal optical depth and temperature.

\begin{table*}[h]
        
        \centering
        \setlength{\tabcolsep}{8.pt}
        \caption{\small{\textit{MoCA} estimates for various AGNs observed by \textit{NuSTAR} characterised by $N_{\rm{H}}$<6$\times 10^{22}$cm$^{-2}$. The first column identifies the source while in columns 2 and 3 the literature values for the high-energy cut-off and the photon index are given. In column number 4 we report the bibliographic references for the phenomenological parameters. Then in columns 5 and 6 (7 and 8) we report the \textit{MoCA} $\tau-kT$ estimates for the slab (sphere) geometry. Positive(negative) errors are estimated by applying our equations on the nominal phenomenological parameter value plus(minus) its associated uncertainty. \label{newtort}}}
        
        \begin{tabular}{c c c c c c c c}
                \hline
        &&&&\multicolumn{2}{c}{Slab}&\multicolumn{2}{c}{Sphere}\\
                Source name & $\Gamma$& E$_{\rm{cut-off}}$ & References &$kT$& $\tau$&$kT$& $\tau$\\
                &&(keV)&&(keV)&&(keV)&\\
                \hline
                \hline
                \\
                \\
Ark 564&2.27$\pm$0.08&42$\pm$3&1&21$^{+2}_{-2}$&2.4$^{+0.4}_{-0.3}$&17$^{+2}_{-1}$&5.0$^{+0.6}_{-0.6}$\\
GRS 1734-292&1.65$\pm$0.05&53$^{+11}_{-8}$&2&23$^{+3}_{-2}$&4.2$^{+0.5}_{-0.6}$&24$^{+3}_{-2}$&6.7$^{+0.7}_{-0.7}$\\
NGC 3783&1.87$\pm$0.04&63$^{+11}_{-8}$&3&$26^{+4}_{-3}$&3.0 $^{+0.5}_{-0.5}$&24$^{+4}_{-3}$&5.3$^{+0.7}_{-0.9}$\\
B1422+231 &1.81$^{+0.07}_{-0.06}$&66$^{+17}_{-12}$& 4&26$^{+8}_{-3}$&3.2$^{+0.6}_{-0.8}$&25$^{+7}_{-4}$&5.4$^{+0.8}_{-1.4}$\\
ESO141-G55&1.94$\pm$0.03&69$^{+14}_{-10}$& 3&30$^{+6}_{-5}$&2.5$^{+0.4}_{-0.4}$&28$^{+8}_{-5}$&4.5$^{+0.8}_{-1.0}$\\
Mrk 348&$1.68\pm0.05$&$79^{+39}_{-19}$&3&31$^{+12}_{-6}$&3.3$^{+0.8}_{-1.0}$&30$^{+15}_{-5}$&5.4$ ^{+1.2}_{-1.8}$\\
4U 1344-60&1.95$\pm$0.03&91$^{+13}_{-10}$&3&39$^{+5}_{-4}$&1.9$^{+0.3}_{-0.3}$&40$^{+6}_{-6}$&3.2$^{+0.5}_{-0.5}$\\     
PG 1247+267&2.33$\pm$0.1&96$^{+130}_{-37}$&5&46$^{+28}_{-14}$&1.0$^{+0.7}_{-0.6}$&51$^{+28}_{-18}$&1.7$^{+1.3}_{-0.9}$\\
Mrk 1040&1.91$\pm0.04$&99$^{+39}_{-22}$&3&41$^{+12}_{-8}$&1.9$^{+0.6}_{-0.6}$&43$^{+14}_{-12}$&3.1$^{+1.2}_{-0.9}$\\\
2MASSJ1614346+470420 &1.98$^{+0.11}_{-0.05}$&106$^{+102}_{-37}$& 4&44$^{+26}_{-11}$&1.6$^{+0.8}_{-0.7}$&48$^{+26}_{-20}$&2.6$^{+1.4}_{-1.2}$\\
NGC 3998&1.79$\pm$0.01&107$^{+27}_{-18}$&6&41$^{+8}_{-6}$&2.2$^{+0.4}_{-0.4}$&43$^{+11}_{-8}$&3.5$^{+0.8}_{-0.7}$\\
SWIFT J2127.4+5654&2.08$\pm$0.01&108$^{+11}_{-10}$&7&46$^{+3}_{-3}$&1.4$^{+0.1}_{-0.1}$&51$^{+4}_{-4}$&2.2$^{+0.2}_{-0.2}$\\
MCG -05-23-16&1.85$\pm$0.01&116$^{+6}_{-5}$&8&45$^{+2}_{-2}$&1.9$^{+0.1}_{-0.1}$&49$^{+2}_{-2}$&3.0$^{+0.2}_{-0.2}$\\
3C390.3&1.71$\pm$0.01&117$^{+18}_{-14}$&9&43$^{+5}_{-4}$&2.3$^{+0.3}_{-0.3}$&46$^{+7}_{-6}$&3.7$^{+0.5}_{-0.5}$\\
Mrk 509&1.78$\pm$0.04&143$^{+72}_{-36}$& 3&52$^{+16}_{-10}$&1.8$^{+0.6}_{-0.6}$&57$^{+18}_{-13}$&2.8$^{+1.0}_{-0.8}$\\         
NGC 6814&1.71$^{+0.04}_{-0.03}$&155$^{+70}_{-35}$&10&54$^{+15}_{-10}$&1.8$^{+0.5}_{-0.6}$&60$^{+17}_{-12}$&2.9$^{+0.9}_{-0.8}$\\
NGC 7469 &1.78$\pm$0.02&170$^{+60}_{-40}$&11&59$^{+12}_{-10}$&1.5$^{+0.4}_{-0.4}$&65$^{+13}_{-12}$&2.4$^{+0.7}_{-0.5}$\\
NGC 4593 &1.69$\pm$0.02&170$^{+160}_{-60}$&12&57$^{+29}_{-17}$&1.8$^{+0.8}_{-0.8}$&63$^{+30}_{-21}$&2.8$^{+1.4}_{-1.1}$\\
MCG +8-11-11&1.77$\pm$0.04&175$^{+110}_{-50}$&10&60$^{+20}_{-13}$&1.5$^{+0.6}_{-0.6}$&66$^{+21}_{-15}$&2.4$^{+0.9}_{-0.8}$\\
Ark 120&1.87$\pm$0.02&183$^{+83}_{-43}$&13&63$^{+14}_{-10}$&1.2$^{+0.3}_{-0.4}$&69$^{+15}_{-11}$&2.0$^{+0.5}_{-0.5}$\\
IC4329A&1.73$\pm$0.01&186$\pm$14&14&62$^{+3}_{-3}$&1.6$^{+0.1}_{-0.1}$&68$^{+3}_{-4}$&2.4$^{+0.2}_{-0.2}$\\             
NGC 4151&1.63$\pm$0.04&196$^{+47}_{-32}$& 15&62$^{+9}_{-8}$&1.8$^{+0.4}_{-0.4}$&70$^{+10}_{-8}$&2.7$^{+0.6}_{-0.5}$\\
3C382&1.68$^{+0.03}_{-0.02}$&214$^{+147}_{-63}$&16&67$^{+23}_{-14}$&1.5$^{+0.5}_{-0.6}$&74$^{+23}_{-15}$&2.4$^{+0.8}_{-0.8}$\\
M81&1.86$\pm$0.01&220$^{+173}_{-86}$&17&70$^{+17}_{-12}$&1.1$^{+0.6}_{-0.4}$&77$^{+23}_{-21}$&1.8$^{+0.8}_{-0.6}$\\

3C 120 &1.87$\pm$0.02&305$^{+142}_{-74}$& 15&83$^{+17}_{-12}$&0.8$^{+0.2}_{-0.3}$&90$^{+15}_{-11}$&1.4$^{+0.3}_{-0.3}$\\
NGC 5506&1.91$\pm0.03$&720$^{+130}_{-190}$&18&123$^{+9}_{-15}$&0.3$^{+0.1}_{-0.1}$&127$^{+8}_{-15}$&0.7$^{+0.2}_{-0.2}$\\
\hline
\hline
\\
MCG 6-30-15&2.061$\pm$0.005&>110&19&>47 &<1.4&>51&<2.2\\
NGC 7213&1.84$\pm$0.03&>140&20&>52 &<1.7&>57&<2.6\\
Mrk 335&2.14$^{+0.02}_{-0.04}$&>174&21&>64 &<0.9&>70&<1.5\\
NGC 2110&1.65$\pm$0.03&>210&22&>66 &<1.7&>73&<2.6\\
Fairall9&1.96$^{+0.01}_{-0.02}$&>240&23&>74 &<0.9&> 81&<1.5\\
ESO 362-G18&1.71$^{+0.03}_{-0.05}$&>241&3&>72&<1.4&>80&<2.2\\
HE 0436-4717 &2.01$\pm0.08$&>280&24&>80 &<0.8&> 86&<1.3\\
NGC 4579&1.81$\pm0.01$&>299&6&>82 &<0.9&> 89&<1.6\\
NGC 2992&1.72$\pm$0.03&>350& 25&>90&<1.0&>96&<1.7\\
Mrk 766&2.22$\pm$0.03&>350&26&>90 &<0.4&>94&<0.8\\
\end{tabular}

\tablebib{
(1)~\cite{Kara17}; (2)~\citet{Tort17}; (3)~ \citet{Rani19}; (4)~\citet{Lanzu19}; (5) \citet{Lanzuisi16}; (6)~\citet{Younes18};
(7)~\citet{Marinucci14a}; (8)~\citet{Balokovic15}; (9)~\citet{Lohfink15}; (10)~ \citet{Tort18a}; (11)~ \citet{Middei18}; 
(12)~\citet{Ursi16}; (13)~\citet{Porquet18}; (14)~\citet{Brenneman14}; (15)~\citet{Fabi17}; (16)~\citet{Ballantyne14}; (17)~\citet{Young18};
(18)~\citet{Matt15}; (19)~\citet{Mari14}; (20)~\citet{ursi15}; (21)~\citet{Parker14};(22)~\citet{Mari15}; (23)~\citet{Lohfink16}; (24)~\citet{Middei18b}; (25)~\citet{Marinucci18m}; (26)~\citet{Risa11,Buisson18}.}
\end{table*}

As expected, the optical depths for the slab-like coronae are systematically smaller than those derived assuming a spherical corona (see Table~\ref{newtort}, Fig.~\ref{smspec} and Fig.~\ref{bagrid}). This is due to the fact that in the spherical corona the optical depth $\tau$ is the radial one, and is therefore the same in all directions, while for the slab-case $\tau$ is the vertical and therefore minimal value. Adopting the equations derived in Sect. 2, we compute contour plots of the sources listed in Table~\ref{newtort}; see bottom panels in Fig.~\ref{bagrid}. When the source high-energy cut-off is well determined, the corresponding contour in the physical parameter space is closed (see purple contours in Fig.~\ref{bagrid}), otherwise the corresponding curves are not limited at high $kT$ (lavender in Fig.~\ref{bagrid}). The average values for the coronal opacity and temperature are $\tau$=1.9$\pm$0.8, $kT$=50$\pm$21 keV and $\tau$=3.1$\pm$1.3, $kT$=53$\pm$23 keV for the slab-like and the spherical geometries, respectively. We obtain a larger $\tau$ and a smaller $kT$ than the ones presented in \cite{Ricci18}, for a slab-like geometry; likely because they included upper limits on the optical depth and lower limits of high-energy cut-off. Moreover, these latter authors computed $kT$ to be E$_{\rm{c}}$/2, thus overestimating the electron temperature, especially at low opacity; see Fig.~\ref{nopop}.\\
\subsection{Comparison with different Comptonisation codes}
\begin{figure*}[h!]
        \includegraphics[width=0.48\textwidth]{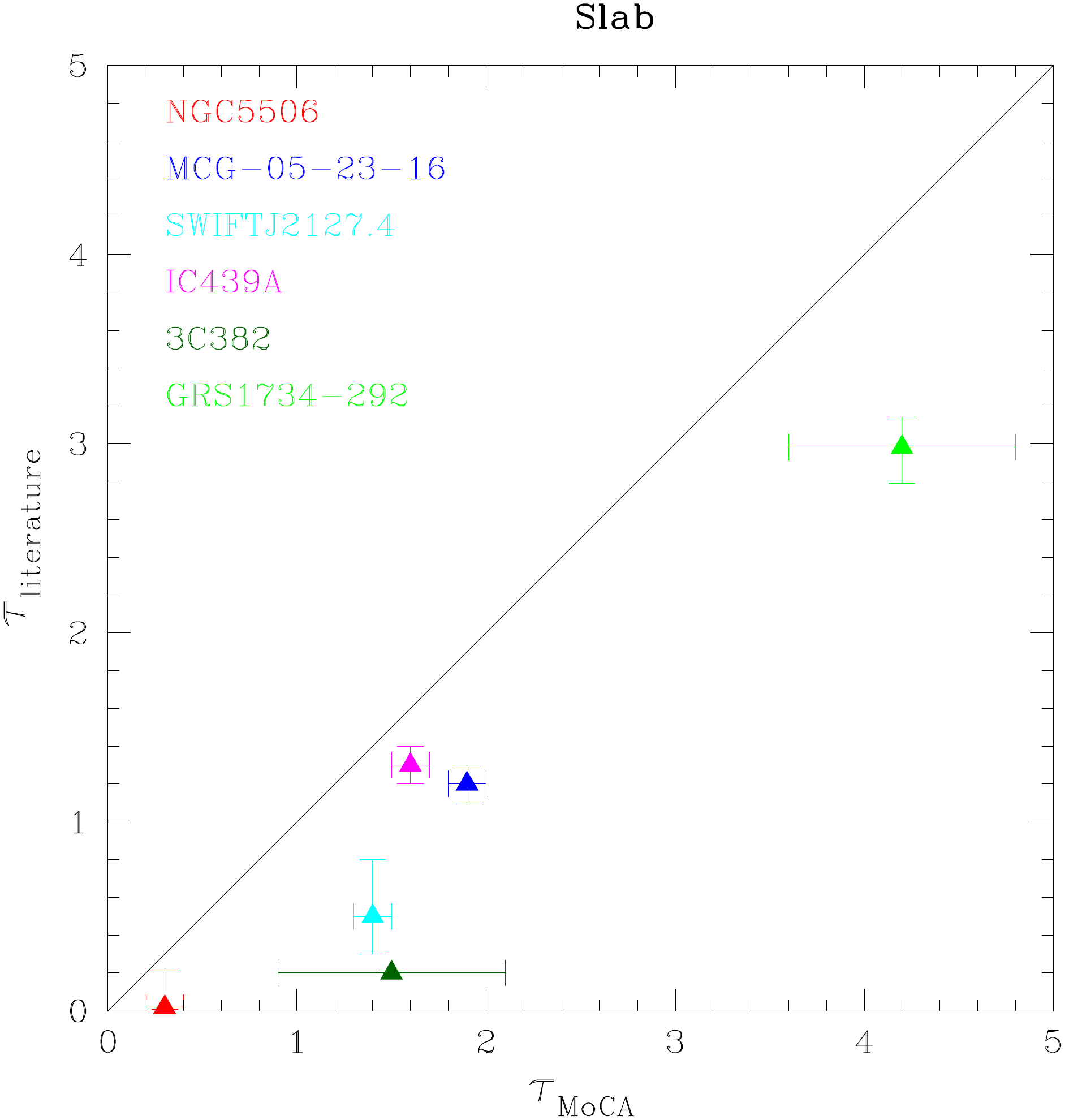}       
        \includegraphics[width=0.48\textwidth]{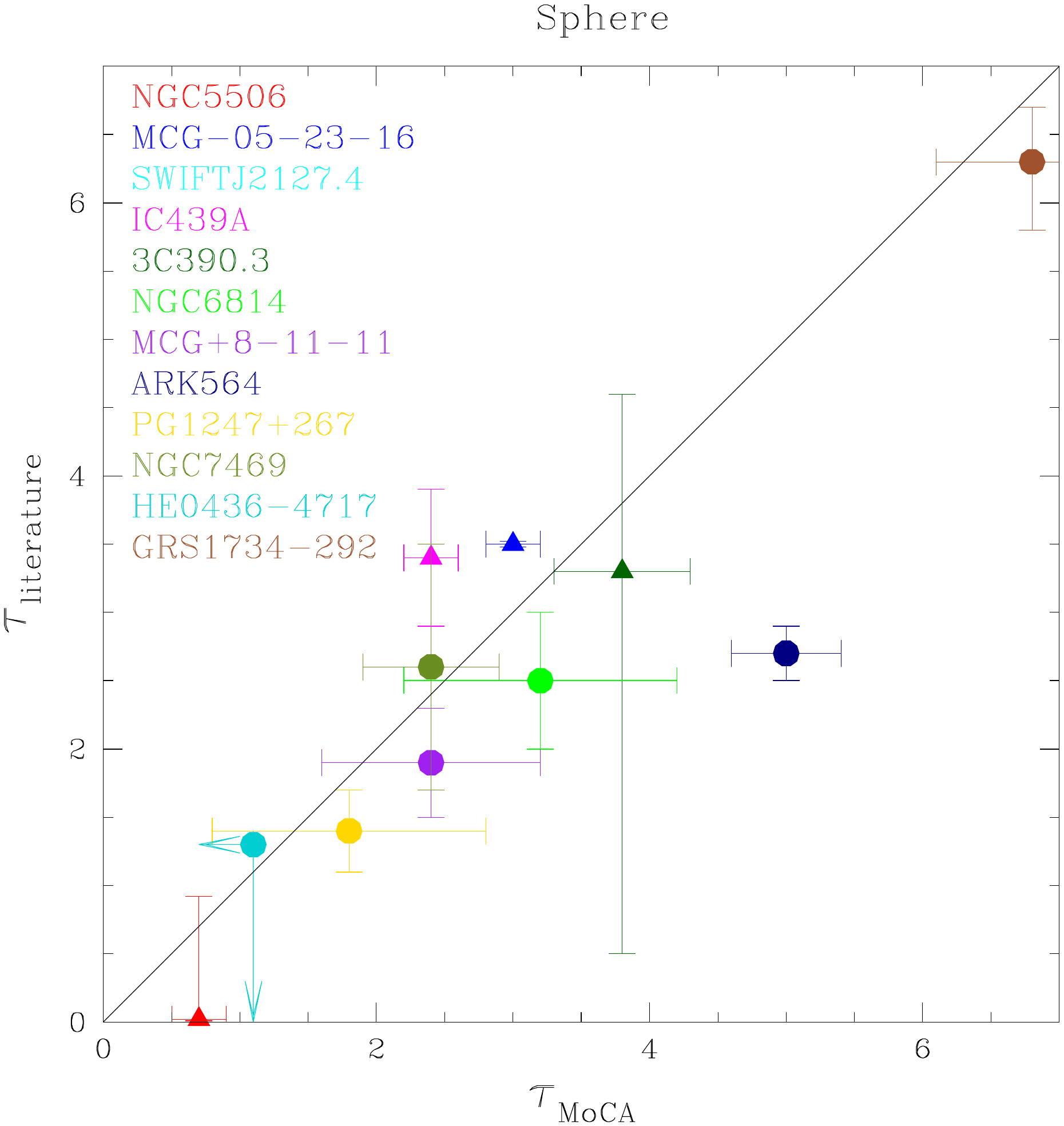}
        \includegraphics[width=0.48\textwidth]{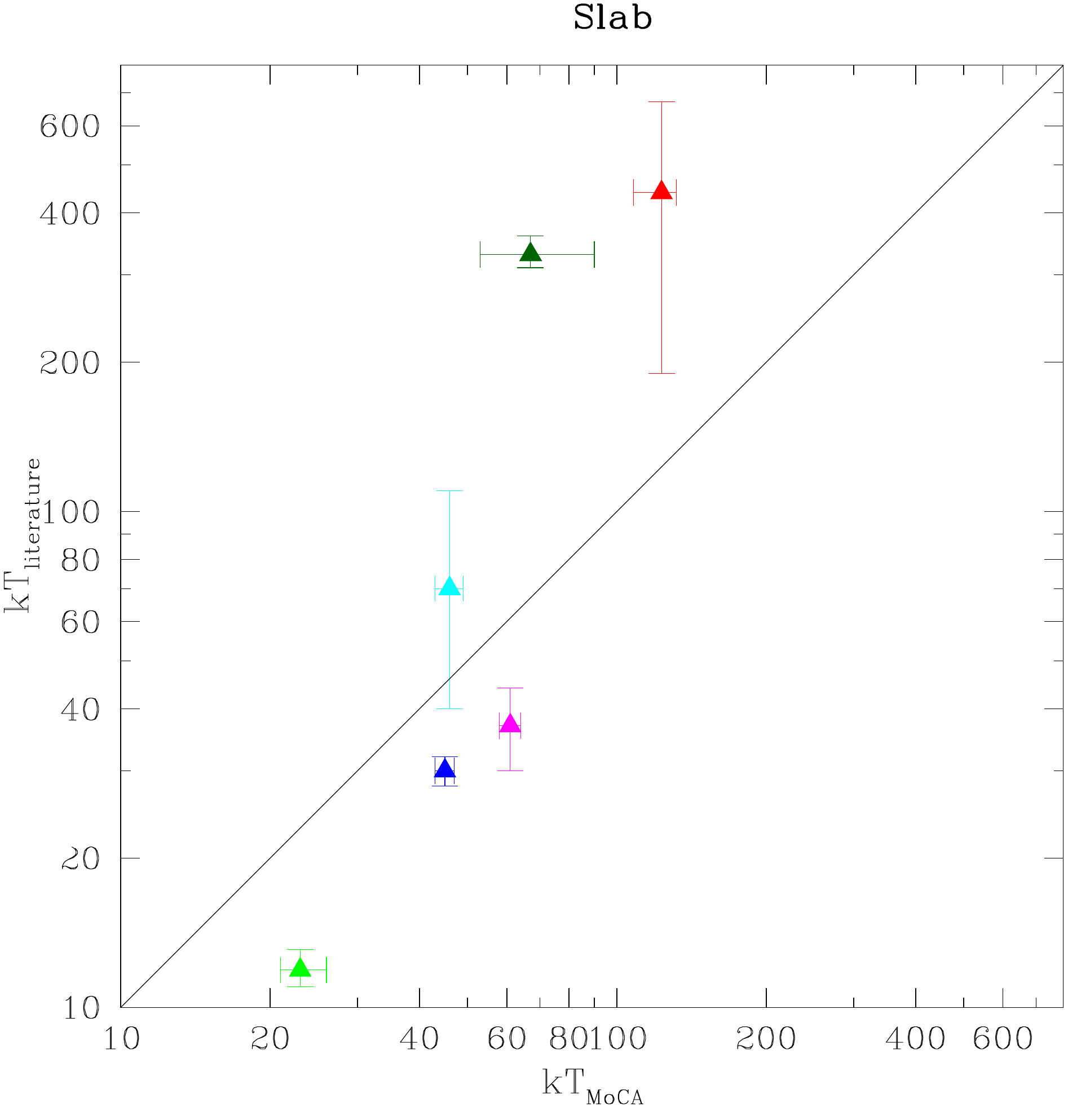}
        \includegraphics[width=0.48\textwidth]{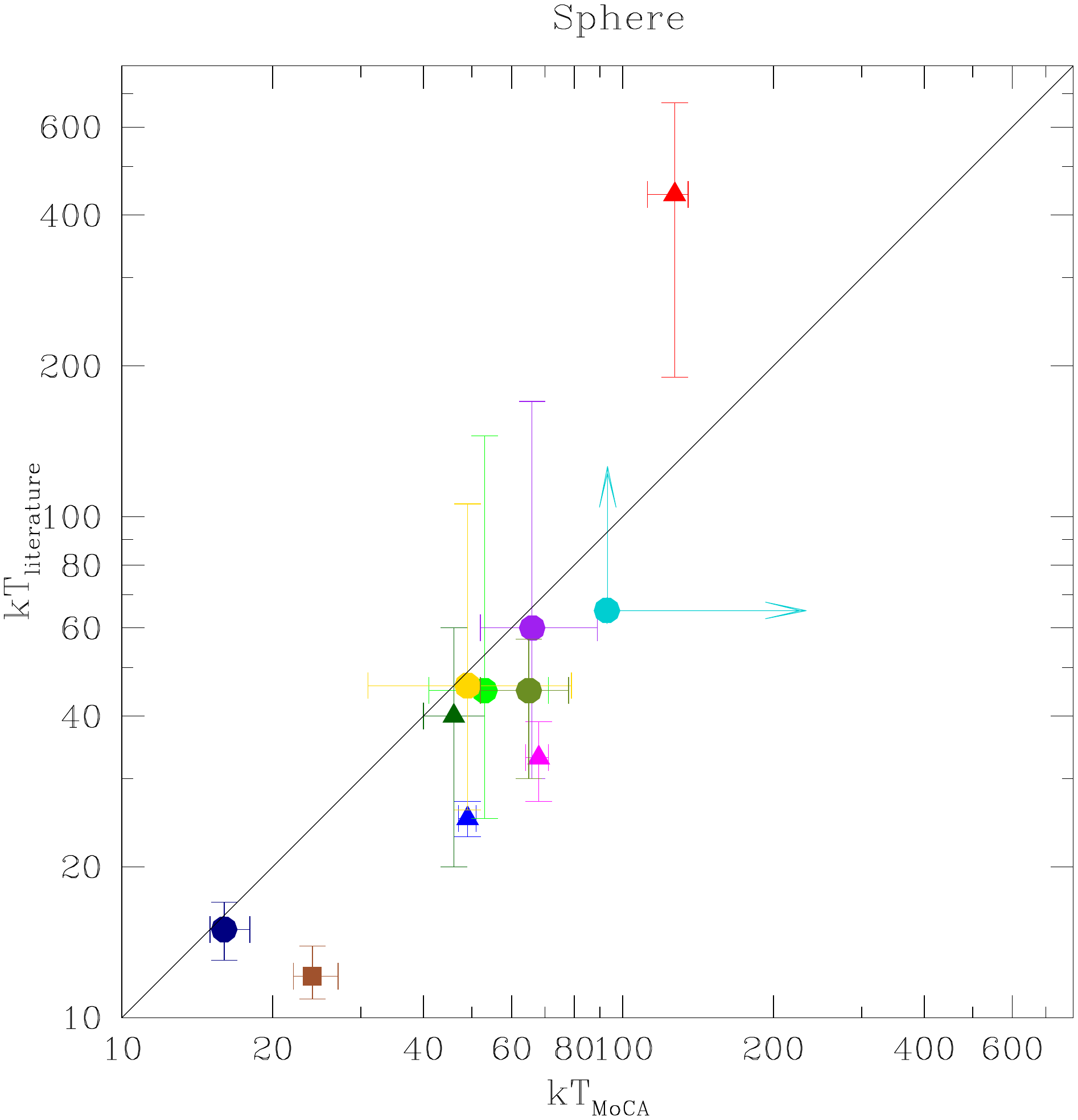}
        \caption{\label{confronti} Comparison of estimates of $kT$ and $\tau$ obtained with MoCA and using different models: \textit{nthcomp} (triangles),  \textit{compTT} (circles), and \textit{compPS} (squares); see \cite{Tort18} and references therein. The 1:1 relation to the graph is also shown. The top and bottom left panels are computed for a slab-like corona, while the remaining panels account for the spherical case. Different coloured names correspond to the various sources that are the same between top and bottom left panels and top and bottom right panels. Temperature and optical depth literature values for the sources are reported by \cite{Tort18}, with the exception of NGC 7469 and HE 0436-4714 for which $kT$ and $\tau$ are reported by \cite{Middei18} and \cite{Middei18b}, respectively.}
\end{figure*}

\indent As a further step, we compared our estimates with those in literature. In particular, in  Fig.~\ref{confronti}, we compare the \textit{MoCA}-based estimates with those reported by other authors and resulting from spectroscopic analyses based on various Comptonisation codes (i.e. \textit{compTT} \citep{Tita94}, \textit{nthcomp} \citep{Zdzi96}, and \textit{compPS} \citep{pout96}). \textit{MoCA}-measurements of the slab-like coronal opacity appear to be larger, on average, than $\tau$ obtained with other models (see top left panel of Fig.~\ref{confronti}), while no trend is found between \textit{MoCA} and literature values of the electron temperature. For the spherical case, \textit{MoCA} estimates of the coronal opacity and electron temperature, except for a few exceptions (e.g. NGC 5506), are compatible with measurements from \textit{compTT}, \textit{nthcomp}, and \textit{compPS}.\\

\section{Summary and Conclusion}

This paper reports on the first systematic application of the code \textit{MoCA} \citep{Tamb18} to study the high-energy emission of AGNs. \textit{MoCA} allowed us to simulate X-ray spectra assuming a wide range of parameters describing the AGNs ($M_{\rm{BH}}$, ${\dot m}$) as well as the Comptonising medium, such as  geometry, temperature, and optical depth. The main results can be summarised as follows:
\begin{itemize}
\item We derived relations between phenomenological ($\Gamma$ and E$_{\rm{c}}$) and physical (electron temperature and optical depth) coronal parameters, assuming either a spherical corona or a slab-like one. We find that $\Gamma$ depends on the Compton parameter $y,$ and also on $kT$. The E$_{\rm{c}}$ is a function of the coronal temperature and the optical depth; see Eqs. \eqref{eqn:fbag},~\eqref{eqn:fsfg},~\eqref{eqn:fbae}, and~\eqref{eqn:fsfe}, and Fig.~\ref{nopop}. Moreover, the relation  E$_{\rm{c}}$=2-3 $kT$ is found to be valid only for a limited range of physical parameters, especially for low $\tau$ and $kT$; see Fig.\ref{nopop}.
\item The derived equations allowed us to determine the loci occupied by the observed AGNs in the physical space of the electron temperature and optical depth (see Fig.~\ref{bagrid}), resulting in a plot useful for deriving the AGN coronal properties from the measured $\Gamma$ and E$_{\rm{c}}$. We then applied the results to an upgraded version of the AGN sample of \cite{Tort18}. We found the impact of the seed photon field spectral shape on the emerging Comptonised spectrum to be weak. The $\Gamma$-E$_{\rm{c}}$ measurements obtained for different black hole masses and accretion rates are consistent within the dispersion with those estimated using Eqs.~\ref{eqn:fbag},~\ref{eqn:fsfg},~\ref{eqn:fbae}, and ~\ref{eqn:fsfe}.
\end{itemize}

\begin{acknowledgements}
We thank the referee for her/his comments.
RM acknowledges Fausto Vagnetti for his precious informatic support, Fondazione Angelo Della Riccia for financial support, Universit\'e Grenoble Alpes and the high energy SHERPAS group for welcoming him at IPAG. Pop acknowledges financial support by the CNES and the Programme National High Energies (PNHE) of CNRS/INSU with INP and IN2P3, co-funded by CEA and CNES. FT acknowledges financial support by CAMK and the project "Modelling the X-ray spectral-timing properties of accreting BH systems with realistic Comptonisation models" (PI: Barbara De Marco).
 
\end{acknowledgements}

\thispagestyle{empty}
\bibliographystyle{aa}
\bibliography{MOCA_accepted_le.bib}

\end{document}